\documentclass[aps,prb,reprint,superscriptaddress,nofootinbib]{revtex4-2}
\usepackage{graphicx,epsfig}
\usepackage{times}
\usepackage{graphics,dcolumn,bm,float}
\usepackage{amssymb,amsmath,rotate,color,dsfont}
\usepackage[breaklinks,colorlinks=true,urlcolor=green,citecolor=green,linkcolor=green]{hyperref}
\usepackage[T1]{fontenc}

\begin{document}
\unitlength 1 cm
\newcommand{\be}{\begin{equation}}
\newcommand{\ee}{\end{equation}}
\newcommand{\bearr}{\begin{eqnarray}}
\newcommand{\eearr}{\end{eqnarray}}
\newcommand{\nn}{\nonumber}
\newcommand{\dagg}{{\dagger}}
\newcommand{\vpdag}{{\vphantom{\dagger}}}
\newcommand{\vecr}{\vec{r}}
\newcommand{\bs}{\boldsymbol}
\newcommand{\up}{\uparrow}
\newcommand{\dn}{\downarrow}
\newcommand{\fns}{\footnotesize}
\newcommand{\ns}{\normalsize}
\newcommand{\cdag}{c^{\dagger}}
\newcommand{\so}{\lambda_{\rm SO}}
\newcommand{\jh}{J_{\rm H}}
\newcommand{\jk}{J_{\rm K}}

\definecolor{red}{rgb}{1.0,0.0,0.0}
\definecolor{green}{rgb}{0.251, 0.502, 0}
\definecolor{blue}{rgb}{0.0,0.0,1.0}

\title{From explicit to spontaneous charge order and the fate of antiferromagnetic quantum Hall state}

\author{Mohsen Hafez-Torbati}
\email{m.hafeztorbati@gmail.com}
\affiliation{Department of Physics, \href{https://ror.org/0091vmj44}{Shahid Beheshti University}, 1983969411, Tehran, Iran}

\begin{abstract}
The antiferromagnetic quantum Hall insulator (AFQHI), where
one of the spin components is in the quantum Hall state and the other in the trivial state, is
an established phase emerging as a result of the Hubbard repulsion in spinful quantum Hall systems.
The stabilization of the AFQHI requires a charge order preventing the effect of the spin-flip
transformation on the electronic state to be compensated by a space-group operation, and is often
induced via an ionic potential. While one would naively expect the nearest-neighbor (NN)
density-density interaction favoring spontaneous charge order to result in qualitatively similar phenomena,
an analysis of the Haldane-Hubbard model extended by the NN interaction finds no
AFQHI. Here, by considering an extended version of the Harper-Hofstadter-Hubbard
model we go beyond the honeycomb structure and suggest that the realization of the AFQHI generically requires an explicit charge
order and cannot be emerged through a spontaneous charge order. We unveil how the AFQHI disappears
upon approaching from the explicit to the spontaneous charge ordering limit.
Our findings shine more light on the stabilization conditions of the AFQHI which
can guide the future optical lattice experiments searching for this intriguing magnetic topological
insulator phase.

\end{abstract}

\maketitle

\section{Introduction}
The Hubbard model serves as a paradigm for the study of the metal to Mott insulator transition \cite{Imada1998}.
Extensions of the model are introduced to search for novel quantum phases of matter that may exist between conventional ones.
Two famous examples are the ionic Hubbard model and the extended Hubbard model.
In the former model, an ionic potential $\Delta$ inducing an explicit charge order is added to the Hubbard Hamiltonian,
while in the latter model a NN interaction $V$ favoring a spontaneous charge order is included.
In one dimension, accurate and consistent results by different methods for both models are available
suggesting a spontaneously dimerized insulator between the quasi-long-range order Mott insulator
at strong and the charge-density-wave insulator (CDWI) at weak Hubbard interaction $U$
\cite{Fabrizio1999,Manmana2004,Tincani2009,HafezTorbati2014,Loida2017,Nakamura2000,Sandvik2004,Ejima2007,HafezTorbati2017a,Spalding2019}.
The main difference between the phase diagrams of the two models is that, in the ionic Hubbard model
the spontaneously dimerized insulator phase persists for large values of
$U$ and $\Delta$ \cite{Manmana2004,Tincani2009} while in the extended Hubbard model it disappears beyond a
critical end-point in the $U$-$V$ phase diagram \cite{Sandvik2004,Ejima2007}.
In two dimensions, the results for the existence and nature of intermediate phase(s) are controversial
\cite{Paris2007,Kancharla2007,Chen2010,HafezTorbati2016,Yao2022,Kundu2024,Althueser2024}
although generally one would expect less tendency towards dimerization and a large tendency towards
long-range AF order in contrast to the one-dimensional case.

Similarly, the Haldane model in the presence of the Hubbard interaction and the ionic potential
is investigated searching for novel topological phases which can be stabilized by strong interaction.
The emergence of an AFQHI with the Chern number $\mathcal{C}=1$, initially reported in a mean-field
theory analysis \cite{He2011}, is confirmed by a variety of methods \cite{Vanhala2016,Tupitsyn2019,Yuan2023,He2024}.
The phase is found to be generic and not restricted only to the honeycomb
structure or to a particular type of the AF or the charge order \cite{Ebrahimkhas2021}.
The essential prerequisite is the absence
of a space-group operation to compensate the effect of the spin-flip transformation on the
electronic state.
Otherwise, the two spin components cannot fall in distinct topological states
and a $\mathcal{C}=1$ AFQHI cannot occur \cite{Ebrahimkhas2021}.
Naively, it might have been expected that replacing the ionic potential with the NN density-density
interaction in the Haldane-Hubbard model would still
support the emergence of the AFQHI phase. At least in the mean-field approximation the NN interaction
reduces to an effective ionic potential conveying the sense that the two terms would exhibit qualitatively
similar phenomena.
However, an analysis of the Haldane-Hubbard model extended by the NN interaction using a combination
of the mean-field theory approximation, density-matrix renormalization group method, and the exact diagonalization (ED)
of finite clusters finds no AFQHI in the phase diagram of the model \cite{Shao2021}.

In this paper, by considering an extended version of the Harper-Hofstadter-Hubbard model we go beyond the honeycomb
structure and suggest that
the absence of the AFQHI in the Haldane-Hubbard model with the NN interaction is not accidental but reflects
a generic feature which goes beyond a specific model.
We confirm that the system either develops a spontaneous charge order, or an AF order, or a non-trivial topology.
The three features never coexist and the AFQHI never stabilizes.
We stabilize the AFQHI through an ionic potential $\Delta$ inducing an explicit charge order in the system.
We map out the phase diagram of the model in the $\Delta$-$V$ plane and show how the
AFQHI disappears beyond a critical end-point as the limit of zero $\Delta$ is approached.
Our findings suggest that the realization of the $\mathcal{C}=1$ AFQHI generically requires an explicit charge order and
cannot be achieved via a spontaneous charge order.

\section{Extended Harper-Hofstadter-Hubbard model}
The Harper-Hofstadter model \cite{Thouless1982} and the Haldane model \cite{Haldane1988}
are two prototype models to acquire Bloch bands
with non-trivial Chern numbers \cite{Haldane2017}.
Both models are realized on optical lattices using the laser-assisted
tunneling \cite{Aidelsburger2013,Miyake2013} and the lattice-shaking techniques \cite{Jotzu2014}.
The high control and tunability of parameters on optical
lattices has motivated the theoretical studies of various extensions of these models
\cite{Juenemann2017,Hofstetter2018,Rachel2018}.
To address the
fundamental question of the role of explicit and spontaneous charge ordering on the stabilization of the AFQHI we
consider the extended Harper-Hofstadter-Hubbard model
\be
H\!=\!H_t+\!\Delta \sum_{\vec{r}}(-1)^{x+y} n^\vpdag_{\vec{r}}
+U\sum_{\vec{r}} n^\vpdag_{\vec{r},\dn} n^\vpdag_{\vec{r},\up}
\!+V\!\!\sum_{\langle \vec{r},\vec{r}' \rangle} \!\! n^\vpdag_{\vec{r}}
n^\vpdag_{\vec{r}'}
\label{eq:ham}
\ee
with the hopping term
\begin{align}
\label{eq:hopping}
\noindent
  & H_t=\!-\sum_{\vec{r},\sigma}
   \left( tc^\dagger_{\vec{r}+\hat{x},\sigma}c^\vpdag_{\vec{r},\sigma}
   +te^{2\pi i \varphi (x+y+\frac{1}{2})} c^\dagger_{\vec{r}+\hat{y},\sigma}c^\vpdag_{\vec{r},\sigma} \right.
   +t'
 \nn \\
&\! \times \!
   \left. e^{2\pi i \varphi (x+y+1)} (c^\dagger_{\vec{r}+\hat{x}+\hat{y},\sigma}
   c^\vpdag_{\vec{r},\sigma}\!+\! c^\dagger_{\vec{r}+\hat{y},\sigma}
   c^\vpdag_{\vec{r}+\hat{x},\sigma})\!+\!{\rm H.c.} \right).
\end{align}
The operators $c^\dag_{\vec{r},\sigma}$ and $c^\vpdag_{\vec{r},\sigma}$ are the fermionic creation
and annihilation operators at the lattice site $\vec{r}$ with the $z$ component of spin $\sigma=\up,\dn$.
The occupation number operator
$n^\vpdag_{\vec{r},\sigma}:=c^\dagger_{\vec{r},\sigma}c^\vpdag_{\vec{r},\sigma}$ and
$n^\vpdag_{\vec{r}}:=n^\vpdag_{\vec{r},\up}+n^\vpdag_{\vec{r},\dn}$.
The summation over $\vec{r}=x\hat{x}+y\hat{y}=(x,y)$ spans the square lattice with the lattice constant set
to unity. We consider the system at half-filling.

\begin{figure}[t]
   \begin{center}
   \includegraphics[width=0.42\textwidth,angle=0]{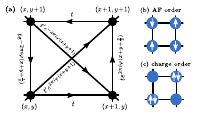}
   \caption{Schematic representation of the hopping term \eqref{eq:hopping} in (a),
   the N\'eel AF order in (b), and the checkerboard
   charge order in (c).}
   \label{fig:ham}
   \end{center}
\end{figure}

The hopping term $H_t$ is the Harper-Hofstadter model extended by the next-nearest-neighbor (NNN) hopping.
The parameter $\varphi$ is the magnetic flux entering the unit square, in units of magnetic flux quantum.
For simplicity we consider
$\varphi=1/2$ which satisfies our purposes here and choose a typical value $t'=0.25t$ for
the NNN hopping throughout this paper.
The presence of the NNN hopping term in Eq. \eqref{eq:hopping} is necessary in order to realize a quantum Hall
state at half-filling otherwise the system would be a semi-metal \cite{Hatsugai1990}.
One notes that the Harper-Hofstadter
model is usually written using the Landau gauge $\vec{A}=Bx\hat{y}$ which makes the hopping phase depend
only on the $x$ coordinate. In writing Eq. \eqref{eq:hopping} we have employed the modified gauge $\vec{A}=B(x+y)\hat{y}$
which in the current problem has the advantage of reducing the number of lattice sites in the unit cell from four to two
in both the checkerboard CDWI phase and in the N\'eel AF phase.
However, we still refer to the model as the Harper-Hofstadter
model since the difference is only a gauge transformation.
The extended Harper-Hofstadter model in Eq. \eqref{eq:hopping}, the N\'eel AF order, and the checkerboard charge order
are schematically depicted in different panels of Fig. \ref{fig:ham}.

The second term in Eq. \eqref{eq:ham} is the ionic potential which gives the onsite energy $+\Delta$
to the even ($A$) sublattice defined by the lattice sites with $x+y$ even and
the onsite energy $-\Delta$
to the odd ($B$) sublattice defined by the lattice sites with $x+y$ odd. The ionic potential induces an explicit
charge order with a larger particle density on the odd and a lower particle density on the even sublattice.
The third term in Eq. \eqref{eq:ham} is the local Hubbard interaction which favours AF ordering. The last
term is the NN interaction.
The notation $\langle \vec{r},\vec{r}' \rangle$ restricts
$\vec{r}$ and $\vec{r}'$ to be NN with the interaction on each lattice bond counted only once.
The NN repulsive interaction favours only one sublattice (even or odd) to be occupied leading to a spontaneous
charge order in the absence of $\Delta$.

One notes that the flux $\varphi$ is due to artificial gauge
fields \cite{Dalibard2011} which is why we have added no Zeeman term to the Hamiltonian in Eq. \eqref{eq:ham}.
In the absence of interaction, $U=V=0$, the Hamiltonian in momentum space reduces to a two-level problem which
represents a quantum Hall insulator (QHI) for $|\Delta|<4t'$ with the Chern number
$\mathcal{C}=\mathcal{C}_\up+\mathcal{C}_\dn=2$ and
a normal insulator for $|\Delta|>4t'$.
The Hamiltonian \eqref{eq:ham} serves as a minimal model to study the role of explicit and spontaneous
charge ordering on the stabilization of the AFQHI.

\section{technical aspects}
To address the Hamiltonian \eqref{eq:ham} in different parameter regimes we employ the dynamical mean-field
theory (DMFT) which is an established technique for strongly correlated systems \cite{Georges1996,Pavarini2022}.
The approach fully takes into
account the local quantum fluctuations but ignores the non-local ones by approximating the self-energy to be
spatially local, $\Sigma_{\vec{r},\vec{r}';\sigma}(i\omega_n)=\Sigma_{\vec{r},\sigma}({i}\omega_n)\delta_{\vec{r},\vec{r}'}$.
The method is extensively applied to different interacting topological systems
\cite{Budich2013,Vanhala2016,Amaricci2018,Irsigler2019,Ebrahimkhas2021,Ebrahimkhas2022,HafezTorbati2024,HafezTorbati2020}.
The phase
diagram of the Haldane-Hubbard model studied by different methods is an example confirming that the results
obtained by DMFT remain qualitatively correct and taking into account the non-local quantum fluctuations can
only slightly modify the phase boundaries \cite{Vanhala2016,Tupitsyn2019,Yuan2023,He2024}.
We opt for the real-space realization of the DMFT \cite{Potthoff1999,Snoek2008} because it permits
access not only to the bulk but also to the edge properties on equal footing. We specifically utilize the
implementation introduced in Ref. \onlinecite{HafezTorbati2018} which we have already successfully applied to various
similar models \cite{Ebrahimkhas2021,Ebrahimkhas2022,HafezTorbati2024,HafezTorbati2020}.
We consider lattices of the size $L\times L$ with $L=40$ and
apply periodic boundary conditions in both directions to analyze the bulk properties. We apply periodic boundary
conditions in $y$ and open boundary conditions in $x$ direction (cylindrical geometry) to analyze the edge properties.
For selective points close to the transition points we have also produced data for $L=60$ corroborating that
the results are independent of the system size. The Anderson impurity problem is solved using the ED method \cite{Caffarel1994}
which provides accurate results for local static quantities and allows direct access to the real-frequency
dynamics \cite{Georges1996}. We mainly consider the number of bath sites $n_b=6$ but for selective points close to the phase
transitions we show that the results remain indistinguishable from the ones obtained for $n_b=5$ and $7$.

In the limit of large coordination number justifying the DMFT approximation the intersite interactions
simplify to their Hartree substitute \cite{MuellerHartmann1989}.
The intersite interactions need to be scaled as $1/D$ because there are $D$ such
interactions per lattice site. This scaling makes only the Hartree energy finite. The Fock and
correlation energies become negligible, of the order of $1/D$ \cite{MuellerHartmann1989}.
We treat the NN interaction in Eq. \eqref{eq:ham}
in the Hartree level,
\begin{align}
\label{eq:hf}
V\!\sum_{\langle \vec{r},\vec{r}' \rangle} \!\! n^\vpdag_{\vec{r}}  n^\vpdag_{\vec{r}'}
\approx
V\langle n^\vpdag_{\!B} \rangle \sum_{\vec{r}\in A} Z^\vpdag_{\vec{r}} n^\vpdag_{\vec{r}}
&+V \langle n^\vpdag_{\!A}\rangle \sum_{\vec{r}\in B} Z^\vpdag_{\vec{r}} n^\vpdag_{\vec{r}}
\nn \\
&-2V \langle n^\vpdag_{\! A}\rangle \langle n^\vpdag_{\! B} \rangle \sum_{\vec{r}} \mathds{1} \ ,
\end{align}
where $\langle n^\vpdag_{A} \rangle$ and $\langle n^\vpdag_{B} \rangle$ denote the particle density
on the sublattice $A$ and on the sublattice $B$, and $Z^\vpdag_{\vec{r}}$ is the coordination
number for the lattice site $\vec{r}$. For periodic boundary conditions in both directions
one simply has $Z^\vpdag_{\vec{r}}=4$ independent of $\vec{r}$. For cylindrical geometries one has
$Z^\vpdag_{\vec{r}}=3$ for the sites at the edges and $Z^\vpdag_{\vec{r}}=4$ for the other sites.
The last contribution written for periodic boundary conditions in both directions
is only relevant when comparing the energies of different solutions in the coexistence regions.

For periodic boundary conditions in both directions, bulk properties, the DMFT loop starts
with an initial guess for the self-energies $\Sigma_{A,\sigma}({i}\omega_n)$ and
$\Sigma_{B,\sigma}({i}\omega_n)$ as well as
the particle densities $\langle n^\vpdag_{A} \rangle$ and $\langle n^\vpdag_{B} \rangle$. The values are updated
at each iteration until the convergence within a prescribed tolerance is reached.
For cylindrical geometry,
we ignore the edge effects on the particle density $\langle n^\vpdag_{\vec{r}} \rangle$, which is already utilized
in writing Eq. \eqref{eq:hf}, and fix the sublattice densities $\langle n^\vpdag_{A} \rangle$ and
$\langle n^\vpdag_{B} \rangle$
to what we have already computed for the bulk case.
This fixes the Hamiltonian and
avoids too many unknown parameters allowing for a faster convergence of the DMFT loop.
The DMFT loop starts with an initial guess for the self-energies $\Sigma_{\vec{r},\sigma}({i}\omega_n)$ of the sites
in the $L\times 2$ unit cell, which are updated at each iteration until the convergence is reached.
One notes that although we ignore the edge effects on the particle densities in the Hartree approximation
in Eq. \eqref{eq:hf}, still the edges effects can be captured via the hopping terms in the Hamiltonian and the
coordination number $Z_{\vec{r}}$ in Eq. \eqref{eq:hf}.
The outcomes for the particle densities at the bulk and at the edges show a maximum difference of about $10\%$,
which quickly drops as the bulk is approached. This confirms that
the edge effects on the particle density is not indeed significant.

To distinguish different phases we compute the staggered charge density $n_s$, the local magnetization
of the N\'eel AF order $m$, and the effective ionic potential $\tilde{\Delta}_\sigma$ given by
\begin{subequations}
\label{eq:param}
 \begin{align}
 \label{eq:ns}
  n_s&=\frac{1}{4}|\langle n^\vpdag_{B} -n^\vpdag_{A} \rangle| \ , \\
  \label{eq:m}
  m&=\frac{1}{2}|\langle n^\vpdag_{\vec{r},\up} -n^\vpdag_{\vec{r},\dn} \rangle| \ , \\
  \label{eq:edel}
\tilde{\Delta}_\sigma&=
\Delta+2V\langle n^\vpdag_{B} -n^\vpdag_{A} \rangle+ \frac{\Sigma_{A,\sigma}(0)-\Sigma_{B,\sigma}(0)}{2} \ .
  \end{align}
\end{subequations}
The staggered charge density $n_s$ is normalized to have the maximum value of $1/2$ similar to
the local magnetization. The effective ionic potential $\tilde{\Delta}_\sigma$ determines if
the spin component $\sigma$ is in the quantum Hall state for $|\tilde{\Delta}_\sigma|<4t'$
or in the normal state for $|\tilde{\Delta}_\sigma|>4t'$.
This characterization of the topological properties of the interacting Hamiltonian \eqref{eq:ham}
relies on the topological Hamiltonian method \cite{Wang2012,Wang2013}.
The method allows the identification of the topological invariant
of an interacting Hamiltonian using an effective non-interacting model, called topological Hamiltonian.
For the current problem, the topological Hamiltonian is given
by the non-interacting part of the Hamiltonian \eqref{eq:ham}, with the ionic potential $\Delta$ substituted
with the effective spin-dependent ionic potential \eqref{eq:edel}.
The spin dependence entering through the self-energies at zero frequency, makes it in principle possible
for the different spin components to fall into distinct topological states and hence allows the AFQHI to stabilize.
Although the topological Hamiltonian approach focuses only on the poles of the Green's function and has some
restrictions in its application \cite{Gurarie2011,Wagner2023,Blason2023}
it can be employed
for the current problem and has already been applied successfully to various similar interacting topological systems
\cite{Budich2013,Vanhala2016,Irsigler2019,Ebrahimkhas2021,HafezTorbati2020}.

Our characterization of the topological nature of different phases does not rely solely on the topological
Hamiltonian method. For a non-trivial topological phase predicted by the topological Hamiltonian method, we affirm
the existence of gapless charge excitations localized at edges directly for the interacting model.
We address the excitations in the bulk and at the edges
using a cylindrical geometry with the open boundary condition in the $x$ direction.
We calculate the single-particle spectral function
\begin{align}
\label{eq:sfx}
 A_x(\omega)&:=\frac{1}{2}\sum_\sigma A_{x,\sigma}(\omega) \nn \\
 &:=\frac{1}{4}\sum_\sigma ( A_{x,y;\sigma}(\omega)+A_{x,y+1;\sigma}(\omega)) \ ,
\end{align}
where $A_{x,y;\sigma}(\omega)$ is the local single-particle spectral function at the position $\vec{r}=x\hat{x}+y\hat{y}$
with spin $\sigma$. Equation \eqref{eq:sfx} involves averaging over the spin and the two
non-equivalent lattice sites in the $y$ direction. In a magnetically ordered phase, the spin-resolved spectral
function $A_{x,\sigma}(\omega)$ allows for the distinction of contributions from different spins to the
spectral function. We use the Lorentzian broadening with a broadening factor of $0.05t$ in the
computation of the spectral function.

It should be mentioned that the cluster DMFT could be employed to go beyond the single-site DMFT
and take into account the non-local quantum fluctuations \cite{Potthoff2018}.
However, the reperiodization scheme used in the cluster DMFT to restore the translational symmetry can
lead to a spurious nonzero Chern number \cite{Gu2019}.
This is a vital drawback for the current study since our main aim is to check if an AF phase with
a nonzero Chern number can emerge.
Hence, we opt for the single-site DMFT.

\section{Results}
While the existence of a $\mathcal{C}=1$ AFQHI is demonstrated by a variety of methods in different systems
showing explicit charge order \cite{He2011,Vanhala2016,Tupitsyn2019,Yuan2023,He2024,Ebrahimkhas2021,Wang2024,Tran2022},
no AFQHI phase is found in the Haldane-Hubbard model extended by the NN interaction \cite{Shao2021}.
We first consider the Hamiltonian \eqref{eq:ham} with no ionic potential, $\Delta=0$, aiming
to unveil whether the absence of the AFQHI in the Haldane-Hubbard model with the NN interaction
is accidental or indicates the fact that the AFQHI cannot generically be stabilized via a spontaneous
charge order and requires an explicit charge order.

\begin{figure}[t]
   \begin{center}
   \includegraphics[width=0.24\textwidth,angle=-90]{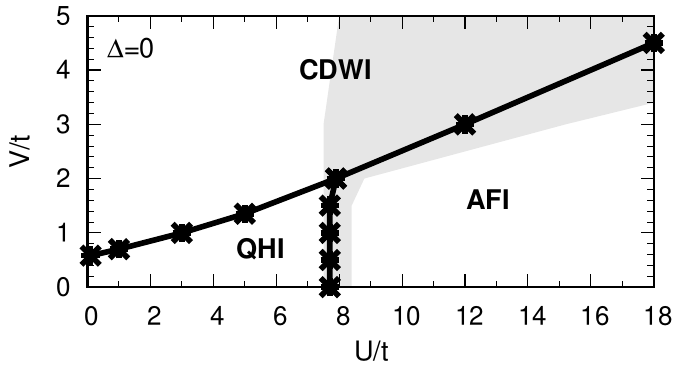}
   \caption{The ground state phase diagram of the Hamiltonian \eqref{eq:ham} for the ionic potential $\Delta=0$.
   The different phases are the quantum Hall insulator (QHI) with the Chern number $\mathcal{C}=2$, the topologically trivial
   N\'eel antiferromagnetic insulator (AFI), and the topologically trivial charge-density-wave insulator (CDWI).
   The gray area indicates the
   coexistence region of the paramagnetic and the AF solutions. The results are for the flux $\varphi=1/2$ and the
   next-nearest-neighbor hopping $t'=0.25t$.}
   \label{fig:pd_d0}
   \end{center}
\end{figure}

Figure \ref{fig:pd_d0} displays
the phase diagram of the model \eqref{eq:ham} for $\Delta=0$ in the plane of the Hubbard $U$ and
the NN interaction $V$.
The results are for the flux $\varphi=1/2$ and the NNN hopping parameter $t'=0.25t$ as for all
the other results in this paper.
Three different phases are identified: The QHI with the Chern
number $\mathcal{C}=\mathcal{C}_\up+\mathcal{C}_\dn=2$ where both $\up$ and $\dn$ spins are in the
quantum Hall state, the topologically trivial N\'eel AF insulator (AFI), and the topologically trivial
CDWI. The gray area specifies the coexistence region of the paramagnetic and the AF solutions.
Transition points within this region are determined by comparing the ground state energies of
the two solutions. The system either has an AF order (AFI), or a charge-density order (CDWI),
or a non-trivial topology (QHI). No AFQHI is found. One notes that a $\mathcal{C}=1$ AFQHI with the
checkerboard charge and the N\'eel AF order could in principle exist; there is no symmetry preventing its
emergence \cite{Ebrahimkhas2021}.

The close similarity between the $U$-$V$ phase diagram \ref{fig:pd_d0} obtained using the
DMFT for the extended Harper-Hofstadter-Hubbard model and the $U$-$V$ phase diagram concluded
in Ref. \onlinecite{Shao2021} for the Haldane-Hubbard model with the NN interaction using a
combination of the mean-field theory, density-matrix renormalization group, and the
ED of finite clusters provides strong evidence that the AFQHI
cannot be stabilized via the NN interaction independent of the details of the model and the lattice structure.
This suggests that the realization of the AFQHI
always requires an explicit charge order and cannot be achieved through a spontaneous charge order.

An AFQHI is expected to exist at a finite $\Delta$ for $V=0$ and is commonly found in the region $U\sim 2\Delta \gg t$
\cite{He2011,Vanhala2016,Ebrahimkhas2021,He2024}. The time-reversal-invariant Hamiltonians are
also reported to host an AF Chern insulator in the same region of the phase diagram \cite{Ebrahimkhas2022}.
A pertinent question that arises and we aim to address next is
how the AFQHI disappears as $V$ is introduced and the limit of zero $\Delta$ is approached.
In the two extreme limits, one can think of an infinitesimal $\Delta$ leading to the emergence
of the AFQHI in the $U$-$V$ phase diagram \ref{fig:pd_d0} or of an infinitesimal $V$ strongly suppressing
the AFQHI in the $U$-$\Delta$ phase diagram.

\begin{figure}[t]
   \begin{center}
   \includegraphics[width=0.24\textwidth,angle=-90]{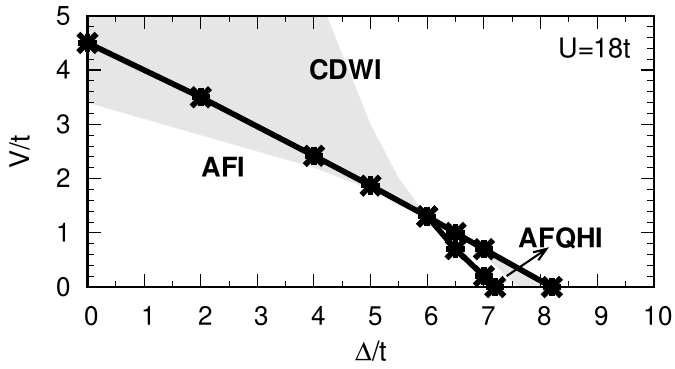}
   \caption{The ground state phase diagram of the Hamiltonian \eqref{eq:ham} for $U=18t$.
   The different phases are the antiferromagnetic quantum Hall insulator (AFQHI) with the Chern number $\mathcal{C}=1$,
   the topologically trivial N\'eel antiferromagnetic insulator (AFI), and the topologically trivial charge-density-wave
   insulator (CDWI). The gray area indicates the coexistence region of the paramagnetic and the AF solutions.
   The results are for the flux $\varphi=1/2$ and the next-nearest-neighbor hopping $t'=0.25t$.}
   \label{fig:pd_U18}
   \end{center}
\end{figure}

Figure \ref{fig:pd_U18} represents the $\Delta$-$V$ phase diagram
of the extended Harper-Hofstadter-Hubbard model \eqref{eq:ham} for the Hubbard interaction $U=18t$.
This relatively large value of the Hubbard interaction is chosen because, according to the previous studies
of similar models, the $\mathcal{C}=1$ AFQHI is expected to stabilize over a wider range of ionic potentials at large
values of $U$ \cite{Ebrahimkhas2021,Ebrahimkhas2022,He2024,Vanhala2016}.
A large Hubbard $U$ is not expected to be an issue for the experimental realization
due to the high-control and tunability of system parameters in optical lattices \cite{Bloch2008}.
Figure \ref{fig:pd_U18} unveils how the AFQHI evolves from the limit of zero $V$ to the limit of zero $\Delta$.
There is always an explicit charge order in the system except for $\Delta \equiv 0$.
For $V=0$ we find that indeed a $\mathcal{C}=1$ AFQHI appears.
The AFQHI is previously reported with the N\'eel AF order on the honeycomb structure
\cite{He2011,Tupitsyn2019,Yuan2023,Vanhala2016,He2024,Wang2024,Tran2022}
and with the stripe AF order on the square lattice \cite{Ebrahimkhas2021}.
The emergence of the AFQHI with the N\'eel AF order on the square lattice in Fig.~\ref{fig:pd_U18} emphasizes
that the AFQHI can exist independent of the type of the (explicit) charge
and AF order, provided the spin-flip symmetry is truly broken \cite{Ebrahimkhas2021}.
One can see from Fig.~\ref{fig:pd_U18} how upon increasing the NN interaction $V$
the AFQHI phase shrinks and disappears at a critical end-point. We expect a similar phase diagram
for the extended Haldane-Hubbard model, explaining the connection between the $U$-$\Delta$ phase diagram
obtained at $V=0$ \cite{Vanhala2016,He2024} and the $U$-$V$ phase diagram obtained at $\Delta=0$ \cite{Shao2021}.
The results in Fig.~\ref{fig:pd_U18} unfold the qualitatively different roles that the ionic potential and the NN interaction
play in the stabilization of the $\mathcal{C}=1$ AFQHI phase.

In the following we present the details of the results leading to the phase diagrams \ref{fig:pd_d0} and \ref{fig:pd_U18}.
We provide data for quantities in Eq. \eqref{eq:param} which allow to characterize the different phases.
In addition, we compare the ground state energies to pinpoint the location of transition points in
coexistence regions. We first set $\Delta=0$ and address Fig. \ref{fig:pd_d0} and then focus on Fig. \ref{fig:pd_U18}.

\begin{figure}[t]
   \begin{center}
   \includegraphics[width=0.58\textwidth,angle=-90]{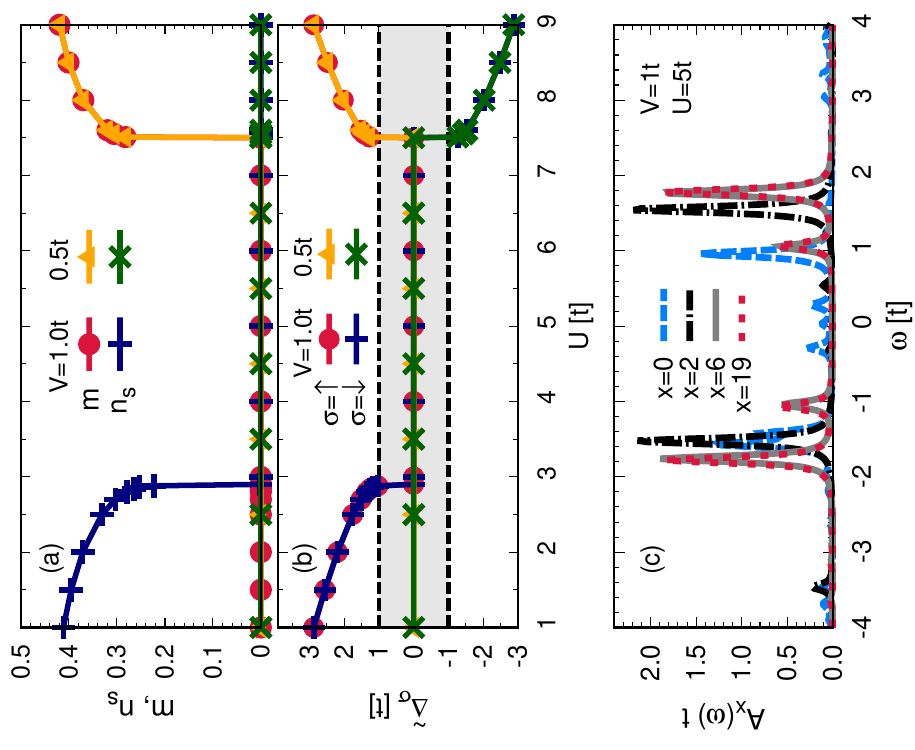}
   \caption{The results for the extended Harper-Hofstadter-Hubbard model \eqref{eq:ham} for
   the ionic potential $\Delta=0$. The staggered charge density $n_s$ and the local magnetization $m$ in panel (a)
   and the effective ionic potential $\tilde{\Delta}_\sigma$ in panel (b) plotted vs the Hubbard interaction $U$ for the
   nearest-neighbor interaction $V=0.5t$ and $1.0t$.
   The dashed lines in (b) separate the topological
   region $|\tilde{\Delta}_\sigma|<4t'$ (shaded area) from the trivial region $|\tilde{\Delta}_\sigma|>4t'$.
   (c) The single-particle spectral
   function \eqref{eq:sfx} vs frequency for $V=1t$ and $U=5t$ at different values of $x$ for a
   cylindrical geometry of the size $40\times40$ with the
   edges at $x=0$ and $x=39$. The data are for the number of bath sites $n_b=6$
   in the ED impurity solver.}
   \label{fig:d0_weakV}
   \end{center}
\end{figure}

Figure \ref{fig:d0_weakV} displays the staggered charge density $n_s$ and the local magnetization $m$ [panel (a)]
and the effective ionic potential $\tilde{\Delta}_\sigma$ [panel (b)] vs the Hubbard interaction $U$
for the two values of the NN interaction $V=0.5t$
and $1.0t$. The results are for $\Delta=0$ and the number of bath sites $n_b=6$ in the ED impurity solver.
The dashed lines in panel (b) separate the topological region (shaded area) from the
trivial region. The spin component $\sigma$ has the Chern number $\mathcal{C}_\sigma=1$
for $|\tilde{\Delta}_\sigma| <4t'$ and the Chern number $\mathcal{C}_\sigma=0$
for $|\tilde{\Delta}_\sigma| >4t'$.
For the small value of the NN interaction $V=0.5t$ there is
no charge-density order in the system in the entire range of the Hubbard $U$.
This involves also $U<t$ not included in the figure. The N\'eel AF phase stabilizes for $U>7.5t$ signaled by a
finite local magnetization $m$. The effective ionic potential for $V=0.5t$ in panel (b)
reveals that the paramagnetic phase is a QHI with the Chern number
$\mathcal{C}=\mathcal{C}_\up+\mathcal{C}_\dn=2$. The effective ionic potential for both $\up$ and $\dn$
spins immediately leaves the topological region as soon as the local magnetization develops indicating
the topologically trivial nature of the AFI.
These results are exemplary for any $V<0.6t$ in the phase diagram \ref{fig:pd_d0}.

In contrast to the results for $V=0.5t$ we find a finite value for the staggered charge density $n_s$ in Fig. \ref{fig:d0_weakV}(a)
for $V=1.0t$ at small values of the Hubbard interaction $U$. The staggered charge density $n_s$ decreases upon increasing $U$
and vanishes at $U \approx 3t$ where the transition from the CDWI to a phase with a uniform charge distribution occurs.
The results for the local magnetization $m$ for $V=1.0t$ in Fig. \ref{fig:d0_weakV}(a) accurately coincide with
the results for $V=0.5t$.
This is why the transition from the QHI to the AFI in Fig. \ref{fig:pd_d0} is almost a straight vertical line.
The effective ionic potential $\tilde{\Delta}_\sigma$ in Fig. \ref{fig:d0_weakV}(b) unveils that the system
either has a local order parameter or a non-trivial topology. As soon as a local order parameter,
either the staggered charge density or the local magnetization, develops, the effective ionic potential for
both spin components leaves the topological region. This confirms that the CDWI and the AFI are topologically
trivial and the paramagnetic phase with the uniform charge distribution is a QHI.

\begin{figure}[t]
   \begin{center}
   \includegraphics[width=0.42\textwidth,angle=-90]{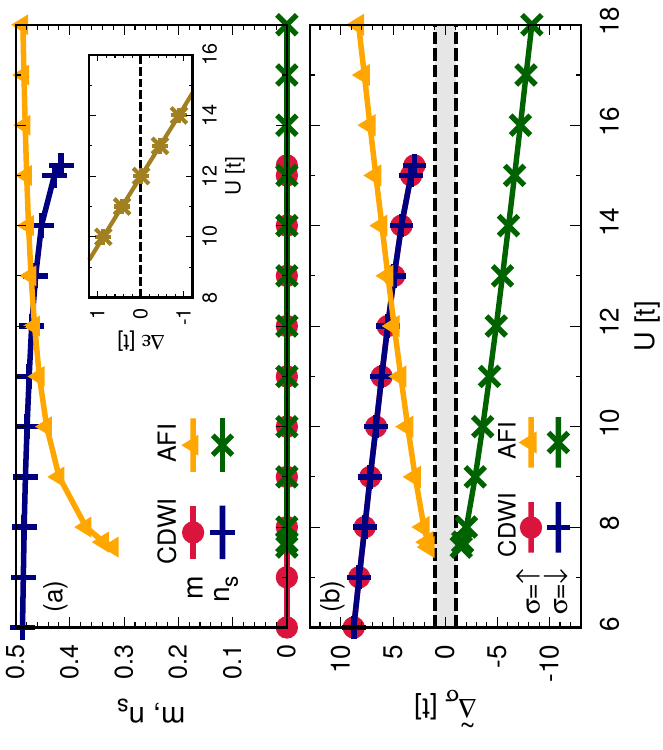}
   \caption{The results for the extended Harper-Hofstadter-Hubbard model \eqref{eq:ham} for
   the ionic potential $\Delta=0$ and the nearest-neighbor  interaction $V=3t$.
   The staggered charge density $n_s$ and the local magnetization $m$ in panel (a)
   and the effective ionic potential $\tilde{\Delta}_\sigma$ in panel (b) are plotted vs the Hubbard interaction $U$
   in the charge-density-wave insulator (CDWI) solution and in the antiferromagnetic insulator (AFI) solution.
   The dashed lines in (b) separate the topological
   region $|\tilde{\Delta}_\sigma|<4t'$ (shaded area) from the trivial region $|\tilde{\Delta}_\sigma|>4t'$.
   The inset represents the ground state energy difference per lattice site of the AFI solution and the CDWI solution,
   $\Delta \varepsilon =\varepsilon^{\vpdag}_{\rm AFI}-\varepsilon^{\vpdag}_{\rm CDWI}$, vs $U$.
   The data are obtained using the number of bath sites $n_b=6$ in the ED impurity solver.}
   \label{fig:d0_V3}
   \end{center}
\end{figure}

Our prediction for the topological nature of the different phases using the effective ionic potential \eqref{eq:edel}
relies on the topological Hamiltonian method which maps the interacting Hamiltonian to an effective non-interacting model.
To characterize the topological properties beyond the topological Hamiltonian method we investigate the charge excitations
in the bulk and at the edges directly for the interacting model using a cylindrical geometry of the size $40\times 40$ with
the edges at $x=0$ and $x=39$. Figure \ref{fig:d0_weakV}(c) shows the single-particle spectral function \eqref{eq:sfx}
vs frequency for the NN interaction $V=1t$ and the Hubbard interaction $U=5t$, where according to the
effective ionic potential $\tilde{\Delta}_\sigma$ in Fig. \ref{fig:d0_weakV}(b) the system is expected to be in the QHI phase.
The results are for the number of bath sites $n_b=6$ in the ED impurity solver.
We find the spectral function $A_x(\omega)$ perfectly symmetric with respect to the center of the cylinder which
is why we have included in Fig. \ref{fig:d0_weakV}(c) only results for $x<20$. One can clearly see that there is a finite
spectral weight at the Fermi energy $\omega=0$ for $x=0$ which quickly vanishes as the bulk is approached.
This certifies the existence of gapless excitations localized at the edges characteristic of the QHI phase.

For large values of the NN interaction $V$ in Fig. \ref{fig:pd_d0} there is a direct transition from the CDWI to the AFI
with a large coexistence region.
To illustrate this portion of the phase diagram we have plotted in Fig. \ref{fig:d0_V3} the staggered charge density $n_s$ and
the local magnetization $m$ [panel (a)] and the effective ionic potential $\tilde{\Delta}_\sigma$ [panel (b)] in both
the CDWI solution and the AFI solution for the NN interaction $V=3t$. One can see from panel (a) that the system either
has a finite local magnetization or a finite staggered charge density.
We have considered different initial guesses for the
self-energies and the electron densities on the $A$ and $B$ sublattices searching for a third possible solution
to the DMFT equations which would exhibit {\it simultaneous} charge and AF order.
However, we did not find such a solution signifying the absence of the AFQHI.

Figure \ref{fig:d0_V3}(b) demonstrates that both the CDWI and the AFI solutions always remain outside the topological
region $|\tilde{\Delta}_\sigma|<4t'$ distinguished by the shaded area.
The difference between the ground state energies per lattice site of the two solutions
$\Delta \varepsilon =\varepsilon^{\vpdag}_{\rm AFI}-\varepsilon^{\vpdag}_{\rm CDWI}$ plotted
vs the Hubbard interaction $U$ in the inset of Fig. \ref{fig:d0_V3} denotes the location
of the transition point from the CDWI to the AFI at $U \simeq 12t$.

For large values of $V$
the system is either in the CDWI with an almost fully polarized staggered charge density $n_s \simeq 0.5$
or in the AFI with an almost fully polarized local magnetization $m \simeq 0.5$. In both cases
the contribution of the hopping term to the ground state energy is expected to be negligible
allowing for the estimates $\varepsilon^{\vpdag}_{\rm CDWI} \simeq U/2$ and
$\varepsilon^{\vpdag}_{\rm AFI} \simeq 2V$. Comparing the two relations one finds the transition occurring
at $V \simeq U/4$. This simple analysis nicely agrees with the results obtained using the DMFT for the transition
from the CDWI to the AFI in Fig. \ref{fig:pd_d0}.

\begin{figure}[t]
   \begin{center}
   \includegraphics[width=0.58\textwidth,angle=-90]{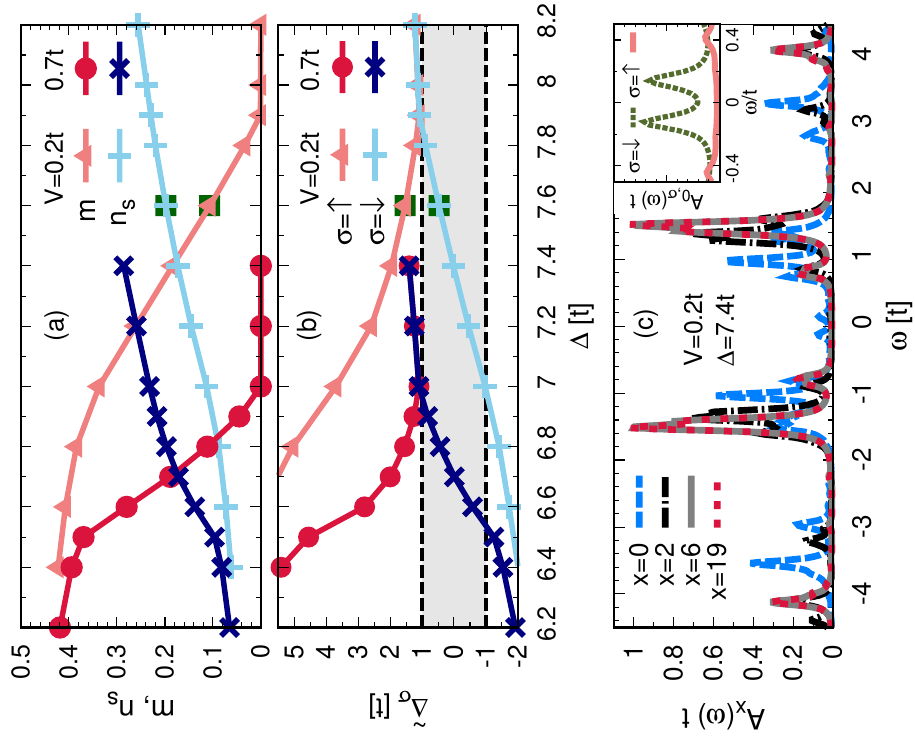}
   \caption{The results for the extended Harper-Hofstadter-Hubbard model \eqref{eq:ham}
   with the fixed Hubbard interaction $U=18t$. The staggered charge density $n_s$ and the
   local magnetization $m$ in panel (a) and the effective ionic potential $\tilde{\Delta}_\sigma$
   in panel (b) plotted vs the ionic potential $\Delta$ for the two different values of the
   nearest-neighbor density-density interaction $V=0.2t$ and $0.7t$. The dashed lines in panel (b)
   separate the topological region $|\tilde{\Delta}_\sigma|<4t'$ (shaded area) from the trivial
   region $|\tilde{\Delta}_\sigma|>4t'$.
   (c) The single-particle spectral function \eqref{eq:sfx} vs frequency for a cylinder of
   the size $40\times40$ with the edges at $x=0$ and $x=39$ for $V=0.2t$ and $\Delta=7.4t$.
   The inset denotes the spin-resolved spectral function $A_{x,\sigma}(\omega)$ at the edge $x=0$.
   The data are for the number of bath sites $n_b=6$ except for the green filled squares
   in panels (a) and (b) at $V=0.2t$ and $\Delta=7.6t$ which are for $n_b=7$.}
   \label{fig:u18_weakv}
   \end{center}
\end{figure}

We now proceed to present results supporting the phase diagram \ref{fig:pd_U18}.
We have plotted in Fig. \ref{fig:u18_weakv} the staggered charge density $n_s$
and the local magnetization $m$ [panel (a)] and the effective ionic potential $\tilde{\Delta}_\sigma$ [panel (b)]
vs the ionic potential $\Delta$ for the two different values of the NN interaction $V=0.2t$ and $0.7t$.
The Hubbard interaction in Fig. \ref{fig:u18_weakv} is fixed to $U=18t$.
One can see in panel (a) a finite staggered charge density which persists
even to small values of $\Delta$ where the system develops long-range AF order.
This is to be compared with the results in Fig. \ref{fig:d0_weakV}(a) where the system
shows either charge or AF order.
The ionic potential inducing a finite staggered charge density in the Mott regime is a point
which is already addressed by both numerical \cite{Manmana2004,Vanhala2016} and analytical \cite{Aligia2004}
calculations.

Figure \ref{fig:u18_weakv}(b) reveals a parameter range where the effective ionic
potential $\tilde{\Delta}_\sigma$ for one spin component, spin $\dn$ in the figure, falls in the
topological region (shaded area) and the other in the trivial region. This confirms the stabilization of
the $\mathcal{C}=1$ AFQHI.
The plotted data are for the AF solution with the positive magnetization on the lower-energy sublattice,
$\langle n^\vpdag_{B,\up}-n^\vpdag_{B,\dn}\rangle >0$.
For the AF solution with the positive magnetization on the higher-energy sublattice $A$, it is the spin
$\up$ which enters the topological region.
One notes how upon increasing the NN interaction $V$
the width of the AFQHI phase decreases.

The data in panels (a) and (b) are obtained with the number of bath sites $n_b=6$ except for the green filled
squares at $V=0.2t$ and $\Delta=7.6t$, which are for $n_b=7$. The results for $n_b=5$ also lie just on
top of the results for $n_b=6$ and $7$.
We have shown the results for different numbers of bath sites at only a single point to avoid overloading the figure.
We have compared the data for the different number of bath sites
$n_b=5$, $6$, and $7$ at multiple other points near the phase boundaries in Figs. \ref{fig:pd_d0} and \ref{fig:pd_U18}
and a similar agreement is found.

Figure \ref{fig:u18_weakv}(c) represents the single-particle spectral function \eqref{eq:sfx}
for a cylinder of the size $40\times 40$ with the edges at $x=0$ and $x=39$ for
the NN interaction $V=0.2t$ and the ionic potential $\Delta=7.4t$,
where according to the results for the effective ionic potential in Fig. \ref{fig:u18_weakv}(b)
the system is expected to be in the $\mathcal{C}=1$ AFQHI phase.
The figure contains only the graphs for $x<20$ because of the perfect symmetry
that the obtained results show with respect to the center of the cylinder.
The number of bath sites $n_b=6$ is used in the ED impurity solver.
There is a finite spectral weight at the Fermi energy $\omega=0$ for
$x=0$ that quickly disappears as the bulk is approached.
The spectral function at $x=6$ already perfectly coincides with the bulk spectral function at $x=19$.
The spin-resolved spectral function $A_{x=0,\sigma}(\omega)$
depicted in the inset of Fig. \ref{fig:u18_weakv}(c) manifests that the finite spectral weight at the
Fermi energy is due to the spin $\dn$ which is the one entering the topological region in Fig. \ref{fig:u18_weakv}(b).
The results verify the existence of gapless excitations localized
at the edges for only one spin component
characteristic of the $\mathcal{C}=1$ AFQHI. The analysis carried out directly for the interacting model
brings additional support for the presence of the AFQHI, independent of the topological Hamiltonian approach.

\begin{figure}[t]
   \begin{center}
   \includegraphics[width=0.41\textwidth,angle=-90]{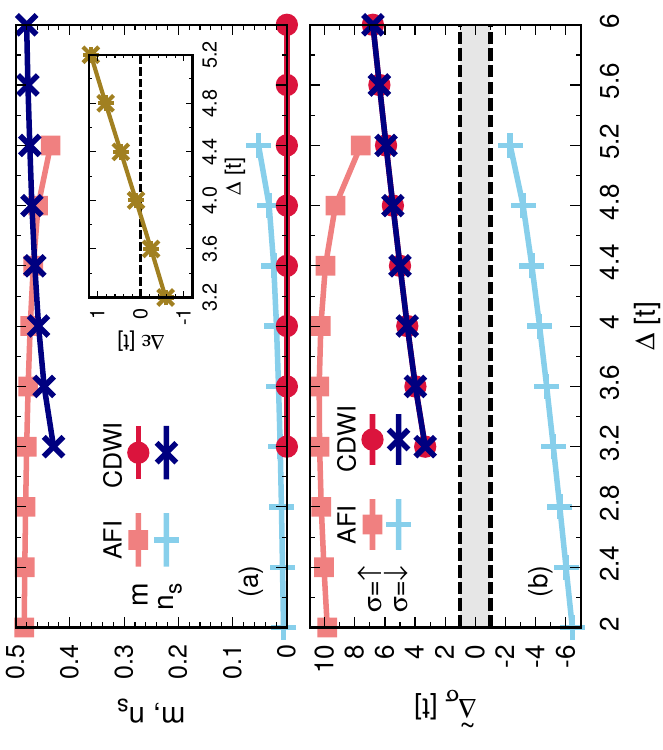}
   \caption{The results for the extended Harper-Hofstadter-Hubbard model \eqref{eq:ham}
   with the Hubbard interaction $U=18t$ and the nearest-neighbor interaction $V=2.5t$.
   The local magnetization $m$ and the staggered charge density $n_s$ in (a) and the effective
   ionic potential $\tilde{\Delta}_\sigma$ in (b) plotted vs the ionic potential $\Delta$ in
   both the antiferromagnetic insulator (AFI) solution and the charge-density-wave insulator (CDWI) solution.
   The dashed lines in (b) separate the topological
   region $|\tilde{\Delta}_\sigma|<4t'$ from the trivial region $|\tilde{\Delta}_\sigma|>4t'$.
   The inset denotes the ground state energy difference per lattice site of the AFI and the CDWI solutions,
   $\Delta \varepsilon=\varepsilon^\vpdag_{\rm AFI}-\varepsilon^\vpdag_{\rm CDWI}$,
   vs $\Delta$. The data are for the number of bath sites $n_b=6$ in the ED
   impurity solver.}
   \label{fig:u18_v2.5}
   \end{center}
\end{figure}

Upon increasing the NN interaction $V$ in Fig. \ref{fig:pd_U18} the intermediate AFQHI phase
disappears and a direct transition from the AFI to the CDWI occurs. This is accompanied by a large coexistence
region of the AFI and the CDWI solutions. For $V=2.5t$ we have represented in
Fig. \ref{fig:u18_v2.5} the staggered charge density $n_s$ and the local magnetization $m$ [panel (a)] and
the effective ionic potential $\tilde{\Delta}_\sigma$ [panel (b)] vs the ionic potential $\Delta$ for both
the AFI solution and the CDWI solution. The two solutions coexist in the region $3.2t \leq \Delta \leq 5.2t$.
The data in panel (b) verify that the effective ionic potential for both
solutions always stays out of the topological region specified by the shaded area. Hence, both solutions always
remain topologically trivial.
The energy difference of the two solutions $\Delta \varepsilon=\varepsilon^\vpdag_{\rm AFI}-\varepsilon^\vpdag_{\rm CDWI}$
plotted in the inset marks the location of the transition from the AFI to the CDWI at $\Delta \simeq 4t$.
We would like to mention that, although we have always presented data for fixed values of $V$, the phase
diagram \ref{fig:pd_U18} is obtained by both fixing $V$ varying $\Delta$,
as well as fixing $\Delta$ varying $V$, showing consistent results for the phase transitions. This has been the case
also for the phase diagram \ref{fig:pd_d0}.

\section{conclusion}
The effect of strong interaction in systems with non-trivial topology can lead to novel quantum
states such as magnetic topological insulators. An example is the spinful quantum Hall systems
subjected to the strong Hubbard interaction hosting the $\mathcal{C}=1$ AFQHI
\cite{He2011,Vanhala2016,Tupitsyn2019,Yuan2023,Ebrahimkhas2021,He2024}.
In this phase, one of the spin
components is in the quantum Hall state and the other one in the trivial state.
In addition to the non-trivial topology and the strong interaction
favouring the AF order, the stabilization of the $\mathcal{C}=1$ AFQHI requires one more ingredient
preventing the effect of the spin-flip transformation on the electronic state to be
compensated by a space-group operation \cite{Ebrahimkhas2021}.
Breaking such a composite symmetry is necessary for the different spin components
to appear in distinct topological states. This is mainly achieved by an ionic potential \cite{He2011,Vanhala2016,Tupitsyn2019,Yuan2023,Ebrahimkhas2021,He2024,Tran2022,Shao2024}
or a sublattice-dependent Hubbard interaction \cite{Wang2024} inducing a charge order in the system.

While the existence of the $\mathcal{C}=1$ AFQHI is already demonstrated in multiple systems with different kinds of charge
and AF order, an analysis of the Haldane-Hubbard model extended by the NN interaction finds no AFQHI phase \cite{Shao2021}.
This is despite the fact that the NN interaction favours a charge order and is expected to support qualitatively
similar phenomena as an ionic potential at least in the mean-field level. This raises the question if the
absence of the $\mathcal{C}=1$ AFQHI in the Haldane-Hubbard model extended by the NN interaction is accidental or reflects
a generic feature.

By considering a minimal extension of the Harper-Hofstadter-Hubbard model we go beyond the honeycomb
structure and study the effect of the ionic potential and the NN interaction on the emergence of the
$\mathcal{C}=1$ AFQHI. In the absence of the ionic potential our results confirm that the system either has an
AF order, or a charge order, or a non-trivial topology. The three features never coexist and
the AFQHI does not emerge. We stabilize the AFQHI through the ionic potential and unveil how the
AFQHI phase shrinks and disappears upon increasing the NN interaction.
Our findings suggest that the emergence of the $\mathcal{C}=1$ AFQHI generically requires an explicit
charge order and cannot be realized via a spontaneous charge order.
This provides more insight into the stabilization conditions of the $\mathcal{C}=1$ AFQHI which paves the path
for future studies searching for this novel magnetic topological state.


\begin{thebibliography}{61}%
\makeatletter
\providecommand \@ifxundefined [1]{%
 \@ifx{#1\undefined}
}%
\providecommand \@ifnum [1]{%
 \ifnum #1\expandafter \@firstoftwo
 \else \expandafter \@secondoftwo
 \fi
}%
\providecommand \@ifx [1]{%
 \ifx #1\expandafter \@firstoftwo
 \else \expandafter \@secondoftwo
 \fi
}%
\providecommand \natexlab [1]{#1}%
\providecommand \enquote  [1]{``#1''}%
\providecommand \bibnamefont  [1]{#1}%
\providecommand \bibfnamefont [1]{#1}%
\providecommand \citenamefont [1]{#1}%
\providecommand \href@noop [0]{\@secondoftwo}%
\providecommand \href [0]{\begingroup \@sanitize@url \@href}%
\providecommand \@href[1]{\@@startlink{#1}\@@href}%
\providecommand \@@href[1]{\endgroup#1\@@endlink}%
\providecommand \@sanitize@url [0]{\catcode `\\12\catcode `\$12\catcode
  `\&12\catcode `\#12\catcode `\^12\catcode `\_12\catcode `\%12\relax}%
\providecommand \@@startlink[1]{}%
\providecommand \@@endlink[0]{}%
\providecommand \url  [0]{\begingroup\@sanitize@url \@url }%
\providecommand \@url [1]{\endgroup\@href {#1}{\urlprefix }}%
\providecommand \urlprefix  [0]{URL }%
\providecommand \Eprint [0]{\href }%
\providecommand \doibase [0]{https://doi.org/}%
\providecommand \selectlanguage [0]{\@gobble}%
\providecommand \bibinfo  [0]{\@secondoftwo}%
\providecommand \bibfield  [0]{\@secondoftwo}%
\providecommand \translation [1]{[#1]}%
\providecommand \BibitemOpen [0]{}%
\providecommand \bibitemStop [0]{}%
\providecommand \bibitemNoStop [0]{.\EOS\space}%
\providecommand \EOS [0]{\spacefactor3000\relax}%
\providecommand \BibitemShut  [1]{\csname bibitem#1\endcsname}%
\let\auto@bib@innerbib\@empty
\bibitem [{\citenamefont {Imada}\ \emph {et~al.}(1998)\citenamefont {Imada},
  \citenamefont {Fujimori},\ and\ \citenamefont {Tokura}}]{Imada1998}%
  \BibitemOpen
  \bibfield  {author} {\bibinfo {author} {\bibfnamefont {M.}~\bibnamefont
  {Imada}}, \bibinfo {author} {\bibfnamefont {A.}~\bibnamefont {Fujimori}},\
  and\ \bibinfo {author} {\bibfnamefont {Y.}~\bibnamefont {Tokura}},\
  }\bibfield  {title} {\bibinfo {title} {Metal-insulator transitions},\ }\href
  {https://doi.org/10.1103/RevModPhys.70.1039} {\bibfield  {journal} {\bibinfo
  {journal} {Rev. Mod. Phys.}\ }\textbf {\bibinfo {volume} {70}},\ \bibinfo
  {pages} {1039–1263} (\bibinfo {year} {1998})}\BibitemShut {NoStop}%
\bibitem [{\citenamefont {Fabrizio}\ \emph {et~al.}(1999)\citenamefont
  {Fabrizio}, \citenamefont {Gogolin},\ and\ \citenamefont
  {Nersesyan}}]{Fabrizio1999}%
  \BibitemOpen
  \bibfield  {author} {\bibinfo {author} {\bibfnamefont {M.}~\bibnamefont
  {Fabrizio}}, \bibinfo {author} {\bibfnamefont {A.~O.}\ \bibnamefont
  {Gogolin}},\ and\ \bibinfo {author} {\bibfnamefont {A.~A.}\ \bibnamefont
  {Nersesyan}},\ }\bibfield  {title} {\bibinfo {title} {From {Band Insulator to
  {Mott} Insulator in One Dimension}},\ }\href
  {https://doi.org/10.1103/PhysRevLett.83.2014} {\bibfield  {journal} {\bibinfo
   {journal} {Phys. Rev. Lett.}\ }\textbf {\bibinfo {volume} {83}},\ \bibinfo
  {pages} {2014–2017} (\bibinfo {year} {1999})}\BibitemShut {NoStop}%
\bibitem [{\citenamefont {Manmana}\ \emph {et~al.}(2004)\citenamefont
  {Manmana}, \citenamefont {Meden}, \citenamefont {Noack},\ and\ \citenamefont
  {Schönhammer}}]{Manmana2004}%
  \BibitemOpen
  \bibfield  {author} {\bibinfo {author} {\bibfnamefont {S.~R.}\ \bibnamefont
  {Manmana}}, \bibinfo {author} {\bibfnamefont {V.}~\bibnamefont {Meden}},
  \bibinfo {author} {\bibfnamefont {R.~M.}\ \bibnamefont {Noack}},\ and\
  \bibinfo {author} {\bibfnamefont {K.}~\bibnamefont {Schönhammer}},\
  }\bibfield  {title} {\bibinfo {title} {Quantum critical behavior of the
  one-dimensional ionic {Hubbard} model},\ }\href
  {https://doi.org/10.1103/PhysRevB.70.155115} {\bibfield  {journal} {\bibinfo
  {journal} {Phys. Rev. B}\ }\textbf {\bibinfo {volume} {70}},\ \bibinfo
  {pages} {155115} (\bibinfo {year} {2004})}\BibitemShut {NoStop}%
\bibitem [{\citenamefont {Tincani}\ \emph {et~al.}(2009)\citenamefont
  {Tincani}, \citenamefont {Noack},\ and\ \citenamefont
  {Baeriswyl}}]{Tincani2009}%
  \BibitemOpen
  \bibfield  {author} {\bibinfo {author} {\bibfnamefont {L.}~\bibnamefont
  {Tincani}}, \bibinfo {author} {\bibfnamefont {R.~M.}\ \bibnamefont {Noack}},\
  and\ \bibinfo {author} {\bibfnamefont {D.}~\bibnamefont {Baeriswyl}},\
  }\bibfield  {title} {\bibinfo {title} {Critical properties of the
  band-insulator-to-{Mott-insulator transition in the strong-coupling limit of
  the ionic Hubbard} model},\ }\href
  {https://doi.org/10.1103/PhysRevB.79.165109} {\bibfield  {journal} {\bibinfo
  {journal} {Phys. Rev. B}\ }\textbf {\bibinfo {volume} {79}},\ \bibinfo
  {pages} {165109} (\bibinfo {year} {2009})}\BibitemShut {NoStop}%
\bibitem [{\citenamefont {{Hafez Torbati}}\ \emph {et~al.}(2014)\citenamefont
  {{Hafez Torbati}}, \citenamefont {Drescher},\ and\ \citenamefont
  {Uhrig}}]{HafezTorbati2014}%
  \BibitemOpen
  \bibfield  {author} {\bibinfo {author} {\bibfnamefont {M.}~\bibnamefont
  {{Hafez Torbati}}}, \bibinfo {author} {\bibfnamefont {N.~A.}\ \bibnamefont
  {Drescher}},\ and\ \bibinfo {author} {\bibfnamefont {G.~S.}\ \bibnamefont
  {Uhrig}},\ }\bibfield  {title} {\bibinfo {title} {Dispersive excitations in
  one-dimensional ionic {Hubbard} model},\ }\href
  {https://doi.org/10.1103/PhysRevB.89.245126} {\bibfield  {journal} {\bibinfo
  {journal} {Phys. Rev. B}\ }\textbf {\bibinfo {volume} {89}},\ \bibinfo
  {pages} {245126} (\bibinfo {year} {2014})}\BibitemShut {NoStop}%
\bibitem [{\citenamefont {Loida}\ \emph {et~al.}(2017)\citenamefont {Loida},
  \citenamefont {Bernier}, \citenamefont {Citro}, \citenamefont {Orignac},\
  and\ \citenamefont {Kollath}}]{Loida2017}%
  \BibitemOpen
  \bibfield  {author} {\bibinfo {author} {\bibfnamefont {K.}~\bibnamefont
  {Loida}}, \bibinfo {author} {\bibfnamefont {J.-S.}\ \bibnamefont {Bernier}},
  \bibinfo {author} {\bibfnamefont {R.}~\bibnamefont {Citro}}, \bibinfo
  {author} {\bibfnamefont {E.}~\bibnamefont {Orignac}},\ and\ \bibinfo {author}
  {\bibfnamefont {C.}~\bibnamefont {Kollath}},\ }\bibfield  {title} {\bibinfo
  {title} {Probing the {Bond Order Wave Phase Transitions of the Ionic
  {Hubbard} Model by Superlattice Modulation Spectroscopy}},\ }\href
  {https://doi.org/10.1103/PhysRevLett.119.230403} {\bibfield  {journal}
  {\bibinfo  {journal} {Phys. Rev. Lett.}\ }\textbf {\bibinfo {volume} {119}},\
  \bibinfo {pages} {230403} (\bibinfo {year} {2017})}\BibitemShut {NoStop}%
\bibitem [{\citenamefont {Nakamura}(2000)}]{Nakamura2000}%
  \BibitemOpen
  \bibfield  {author} {\bibinfo {author} {\bibfnamefont {M.}~\bibnamefont
  {Nakamura}},\ }\bibfield  {title} {\bibinfo {title} {Tricritical behavior in
  the extended {Hubbard} chains},\ }\href
  {https://doi.org/10.1103/PhysRevB.61.16377} {\bibfield  {journal} {\bibinfo
  {journal} {Phys. Rev. B}\ }\textbf {\bibinfo {volume} {61}},\ \bibinfo
  {pages} {16377–16392} (\bibinfo {year} {2000})}\BibitemShut {NoStop}%
\bibitem [{\citenamefont {Sandvik}\ \emph {et~al.}(2004)\citenamefont
  {Sandvik}, \citenamefont {Balents},\ and\ \citenamefont
  {Campbell}}]{Sandvik2004}%
  \BibitemOpen
  \bibfield  {author} {\bibinfo {author} {\bibfnamefont {A.~W.}\ \bibnamefont
  {Sandvik}}, \bibinfo {author} {\bibfnamefont {L.}~\bibnamefont {Balents}},\
  and\ \bibinfo {author} {\bibfnamefont {D.~K.}\ \bibnamefont {Campbell}},\
  }\bibfield  {title} {\bibinfo {title} {Ground {State Phases of the
  Half-Filled One-Dimensional Extended Hubbard Model}},\ }\href
  {https://doi.org/10.1103/PhysRevLett.92.236401} {\bibfield  {journal}
  {\bibinfo  {journal} {Phys. Rev. Lett.}\ }\textbf {\bibinfo {volume} {92}},\
  \bibinfo {pages} {236401} (\bibinfo {year} {2004})}\BibitemShut {NoStop}%
\bibitem [{\citenamefont {Ejima}\ and\ \citenamefont
  {Nishimoto}(2007)}]{Ejima2007}%
  \BibitemOpen
  \bibfield  {author} {\bibinfo {author} {\bibfnamefont {S.}~\bibnamefont
  {Ejima}}\ and\ \bibinfo {author} {\bibfnamefont {S.}~\bibnamefont
  {Nishimoto}},\ }\bibfield  {title} {\bibinfo {title} {Phase {Diagram of the
  One-Dimensional Half-Filled Extended Hubbard Model}},\ }\href
  {https://doi.org/10.1103/PhysRevLett.99.216403} {\bibfield  {journal}
  {\bibinfo  {journal} {Phys. Rev. Lett.}\ }\textbf {\bibinfo {volume} {99}},\
  \bibinfo {pages} {216403} (\bibinfo {year} {2007})}\BibitemShut {NoStop}%
\bibitem [{\citenamefont {Hafez-Torbati}\ and\ \citenamefont
  {Uhrig}(2017)}]{HafezTorbati2017a}%
  \BibitemOpen
  \bibfield  {author} {\bibinfo {author} {\bibfnamefont {M.}~\bibnamefont
  {Hafez-Torbati}}\ and\ \bibinfo {author} {\bibfnamefont {G.~S.}\ \bibnamefont
  {Uhrig}},\ }\bibfield  {title} {\bibinfo {title} {Singlet exciton
  condensation and bond-order-wave phase in the extended {Hubbard} model},\
  }\href {https://doi.org/10.1103/PhysRevB.96.125129} {\bibfield  {journal}
  {\bibinfo  {journal} {Phys. Rev. B}\ }\textbf {\bibinfo {volume} {96}},\
  \bibinfo {pages} {125129} (\bibinfo {year} {2017})}\BibitemShut {NoStop}%
\bibitem [{\citenamefont {Spalding}\ \emph {et~al.}(2019)\citenamefont
  {Spalding}, \citenamefont {Tsai},\ and\ \citenamefont
  {Campbell}}]{Spalding2019}%
  \BibitemOpen
  \bibfield  {author} {\bibinfo {author} {\bibfnamefont {J.}~\bibnamefont
  {Spalding}}, \bibinfo {author} {\bibfnamefont {S.-W.}\ \bibnamefont {Tsai}},\
  and\ \bibinfo {author} {\bibfnamefont {D.~K.}\ \bibnamefont {Campbell}},\
  }\bibfield  {title} {\bibinfo {title} {Critical entanglement for the
  half-filled extended hubbard model},\ }\href
  {https://doi.org/10.1103/PhysRevB.99.195445} {\bibfield  {journal} {\bibinfo
  {journal} {Phys. Rev. B}\ }\textbf {\bibinfo {volume} {99}},\ \bibinfo
  {pages} {195445} (\bibinfo {year} {2019})}\BibitemShut {NoStop}%
\bibitem [{\citenamefont {Paris}\ \emph {et~al.}(2007)\citenamefont {Paris},
  \citenamefont {Bouadim}, \citenamefont {Hébert}, \citenamefont {Batrouni},\
  and\ \citenamefont {Scalettar}}]{Paris2007}%
  \BibitemOpen
  \bibfield  {author} {\bibinfo {author} {\bibfnamefont {N.}~\bibnamefont
  {Paris}}, \bibinfo {author} {\bibfnamefont {K.}~\bibnamefont {Bouadim}},
  \bibinfo {author} {\bibfnamefont {F.}~\bibnamefont {Hébert}}, \bibinfo
  {author} {\bibfnamefont {G.~G.}\ \bibnamefont {Batrouni}},\ and\ \bibinfo
  {author} {\bibfnamefont {R.~T.}\ \bibnamefont {Scalettar}},\ }\bibfield
  {title} {\bibinfo {title} {Quantum {M}onte~{Carlo Study of an
  Interaction-Driven Band-Insulator-to-Metal Transition}},\ }\href
  {https://doi.org/10.1103/PhysRevLett.98.046403} {\bibfield  {journal}
  {\bibinfo  {journal} {Phys. Rev. Lett.}\ }\textbf {\bibinfo {volume} {98}},\
  \bibinfo {pages} {046403} (\bibinfo {year} {2007})}\BibitemShut {NoStop}%
\bibitem [{\citenamefont {Kancharla}\ and\ \citenamefont
  {Dagotto}(2007)}]{Kancharla2007}%
  \BibitemOpen
  \bibfield  {author} {\bibinfo {author} {\bibfnamefont {S.~S.}\ \bibnamefont
  {Kancharla}}\ and\ \bibinfo {author} {\bibfnamefont {E.}~\bibnamefont
  {Dagotto}},\ }\bibfield  {title} {\bibinfo {title} {Correlated {Insulated
  Phase Suggests Bond Order between Band and {M}ott Insulators in Two
  Dimensions}},\ }\href {https://doi.org/10.1103/PhysRevLett.98.016402}
  {\bibfield  {journal} {\bibinfo  {journal} {Phys. Rev. Lett.}\ }\textbf
  {\bibinfo {volume} {98}},\ \bibinfo {pages} {016402} (\bibinfo {year}
  {2007})}\BibitemShut {NoStop}%
\bibitem [{\citenamefont {Chen}\ \emph {et~al.}(2010)\citenamefont {Chen},
  \citenamefont {Zhao}, \citenamefont {Lin},\ and\ \citenamefont
  {Wu}}]{Chen2010}%
  \BibitemOpen
  \bibfield  {author} {\bibinfo {author} {\bibfnamefont {H.-M.}\ \bibnamefont
  {Chen}}, \bibinfo {author} {\bibfnamefont {H.}~\bibnamefont {Zhao}}, \bibinfo
  {author} {\bibfnamefont {H.-Q.}\ \bibnamefont {Lin}},\ and\ \bibinfo {author}
  {\bibfnamefont {C.-Q.}\ \bibnamefont {Wu}},\ }\bibfield  {title} {\bibinfo
  {title} {Bond-located spin density wave phase in the two-dimensional {(2D)
  ionic Hubbard} model},\ }\href
  {http://stacks.iop.org/1367-2630/12/i=9/a=093021} {\bibfield  {journal}
  {\bibinfo  {journal} {New Journal of Physics}\ }\textbf {\bibinfo {volume}
  {12}},\ \bibinfo {pages} {093021} (\bibinfo {year} {2010})}\BibitemShut
  {NoStop}%
\bibitem [{\citenamefont {Hafez-Torbati}\ and\ \citenamefont
  {Uhrig}(2016)}]{HafezTorbati2016}%
  \BibitemOpen
  \bibfield  {author} {\bibinfo {author} {\bibfnamefont {M.}~\bibnamefont
  {Hafez-Torbati}}\ and\ \bibinfo {author} {\bibfnamefont {G.~S.}\ \bibnamefont
  {Uhrig}},\ }\bibfield  {title} {\bibinfo {title} {Orientational bond and
  {N}\'eel order in the two-dimensional ionic {H}ubbard model},\ }\href
  {https://doi.org/10.1103/PhysRevB.93.195128} {\bibfield  {journal} {\bibinfo
  {journal} {Phys. Rev. B}\ }\textbf {\bibinfo {volume} {93}},\ \bibinfo
  {pages} {195128} (\bibinfo {year} {2016})}\BibitemShut {NoStop}%
\bibitem [{\citenamefont {Yao}\ \emph {et~al.}(2022)\citenamefont {Yao},
  \citenamefont {Wang},\ and\ \citenamefont {Wang}}]{Yao2022}%
  \BibitemOpen
  \bibfield  {author} {\bibinfo {author} {\bibfnamefont {M.}~\bibnamefont
  {Yao}}, \bibinfo {author} {\bibfnamefont {D.}~\bibnamefont {Wang}},\ and\
  \bibinfo {author} {\bibfnamefont {Q.-H.}\ \bibnamefont {Wang}},\ }\bibfield
  {title} {\bibinfo {title} {Determinant quantum monte carlo for the
  half-filled {H}ubbard model with nonlocal density-density interactions},\
  }\href {https://doi.org/10.1103/PhysRevB.106.195121} {\bibfield  {journal}
  {\bibinfo  {journal} {Phys. Rev. B}\ }\textbf {\bibinfo {volume} {106}},\
  \bibinfo {pages} {195121} (\bibinfo {year} {2022})}\BibitemShut {NoStop}%
\bibitem [{\citenamefont {Kundu}\ and\ \citenamefont
  {Sénéchal}(2024)}]{Kundu2024}%
  \BibitemOpen
  \bibfield  {author} {\bibinfo {author} {\bibfnamefont {S.}~\bibnamefont
  {Kundu}}\ and\ \bibinfo {author} {\bibfnamefont {D.}~\bibnamefont
  {Sénéchal}},\ }\bibfield  {title} {\bibinfo {title} {{CDMFT+HFD: An
  extension of dynamical mean field theory for nonlocal interactions applied to
  the single band extended Hubbard model}},\ }\href
  {https://doi.org/10.21468/SciPostPhysCore.7.2.033} {\bibfield  {journal}
  {\bibinfo  {journal} {SciPost Phys. Core}\ }\textbf {\bibinfo {volume} {7}},\
  \bibinfo {pages} {033} (\bibinfo {year} {2024})}\BibitemShut {NoStop}%
\bibitem [{\citenamefont {Alth\"user}\ and\ \citenamefont
  {Uhrig}(2024)}]{Althueser2024}%
  \BibitemOpen
  \bibfield  {author} {\bibinfo {author} {\bibfnamefont {J.}~\bibnamefont
  {Alth\"user}}\ and\ \bibinfo {author} {\bibfnamefont {G.~S.}\ \bibnamefont
  {Uhrig}},\ }\bibfield  {title} {\bibinfo {title} {Collective excitations in
  competing phases in two and three dimensions},\ }\href
  {https://doi.org/10.1103/PhysRevB.109.205153} {\bibfield  {journal} {\bibinfo
   {journal} {Phys. Rev. B}\ }\textbf {\bibinfo {volume} {109}},\ \bibinfo
  {pages} {205153} (\bibinfo {year} {2024})}\BibitemShut {NoStop}%
\bibitem [{\citenamefont {He}\ \emph {et~al.}(2011)\citenamefont {He},
  \citenamefont {Zong}, \citenamefont {Kou}, \citenamefont {Liang},\ and\
  \citenamefont {Feng}}]{He2011}%
  \BibitemOpen
  \bibfield  {author} {\bibinfo {author} {\bibfnamefont {J.}~\bibnamefont
  {He}}, \bibinfo {author} {\bibfnamefont {Y.-H.}\ \bibnamefont {Zong}},
  \bibinfo {author} {\bibfnamefont {S.-P.}\ \bibnamefont {Kou}}, \bibinfo
  {author} {\bibfnamefont {Y.}~\bibnamefont {Liang}},\ and\ \bibinfo {author}
  {\bibfnamefont {S.}~\bibnamefont {Feng}},\ }\bibfield  {title} {\bibinfo
  {title} {Topological spin density waves in the {H}ubbard model on a honeycomb
  lattice},\ }\href {https://link.aps.org/doi/10.1103/PhysRevB.84.035127}
  {\bibfield  {journal} {\bibinfo  {journal} {Phys. Rev. B}\ }\textbf {\bibinfo
  {volume} {84}},\ \bibinfo {pages} {035127} (\bibinfo {year}
  {2011})}\BibitemShut {NoStop}%
\bibitem [{\citenamefont {Vanhala}\ \emph {et~al.}(2016)\citenamefont
  {Vanhala}, \citenamefont {Siro}, \citenamefont {Liang}, \citenamefont
  {Troyer}, \citenamefont {Harju},\ and\ \citenamefont
  {Törmä}}]{Vanhala2016}%
  \BibitemOpen
  \bibfield  {author} {\bibinfo {author} {\bibfnamefont {T.~I.}\ \bibnamefont
  {Vanhala}}, \bibinfo {author} {\bibfnamefont {T.}~\bibnamefont {Siro}},
  \bibinfo {author} {\bibfnamefont {L.}~\bibnamefont {Liang}}, \bibinfo
  {author} {\bibfnamefont {M.}~\bibnamefont {Troyer}}, \bibinfo {author}
  {\bibfnamefont {A.}~\bibnamefont {Harju}},\ and\ \bibinfo {author}
  {\bibfnamefont {P.}~\bibnamefont {Törmä}},\ }\bibfield  {title} {\bibinfo
  {title} {Topological {Phase Transitions in the Repulsively Interacting
  {H}aldane-{H}ubbard Model}},\ }\href
  {https://doi.org/10.1103/PhysRevLett.116.225305} {\bibfield  {journal}
  {\bibinfo  {journal} {Phys. Rev. Lett.}\ }\textbf {\bibinfo {volume} {116}},\
  \bibinfo {pages} {225305} (\bibinfo {year} {2016})}\BibitemShut {NoStop}%
\bibitem [{\citenamefont {Tupitsyn}\ and\ \citenamefont
  {Prokof'ev}(2019)}]{Tupitsyn2019}%
  \BibitemOpen
  \bibfield  {author} {\bibinfo {author} {\bibfnamefont {I.~S.}\ \bibnamefont
  {Tupitsyn}}\ and\ \bibinfo {author} {\bibfnamefont {N.~V.}\ \bibnamefont
  {Prokof'ev}},\ }\bibfield  {title} {\bibinfo {title} {Phase diagram topology
  of the {Haldane-Hubbard-Coulomb} model},\ }\href
  {https://doi.org/10.1103/PhysRevB.99.121113} {\bibfield  {journal} {\bibinfo
  {journal} {Phys. Rev. B}\ }\textbf {\bibinfo {volume} {99}},\ \bibinfo
  {pages} {121113} (\bibinfo {year} {2019})}\BibitemShut {NoStop}%
\bibitem [{\citenamefont {Yuan}\ \emph {et~al.}(2023)\citenamefont {Yuan},
  \citenamefont {Guo}, \citenamefont {Lu}, \citenamefont {Lu},\ and\
  \citenamefont {Shao}}]{Yuan2023}%
  \BibitemOpen
  \bibfield  {author} {\bibinfo {author} {\bibfnamefont {H.}~\bibnamefont
  {Yuan}}, \bibinfo {author} {\bibfnamefont {Y.}~\bibnamefont {Guo}}, \bibinfo
  {author} {\bibfnamefont {R.}~\bibnamefont {Lu}}, \bibinfo {author}
  {\bibfnamefont {H.}~\bibnamefont {Lu}},\ and\ \bibinfo {author}
  {\bibfnamefont {C.}~\bibnamefont {Shao}},\ }\bibfield  {title} {\bibinfo
  {title} {Phase transitions in the {H}aldane-{H}ubbard model with ionic
  potentials},\ }\href {https://doi.org/10.1103/PhysRevB.107.075150} {\bibfield
   {journal} {\bibinfo  {journal} {Phys. Rev. B}\ }\textbf {\bibinfo {volume}
  {107}},\ \bibinfo {pages} {075150} (\bibinfo {year} {2023})}\BibitemShut
  {NoStop}%
\bibitem [{\citenamefont {He}\ \emph {et~al.}(2024)\citenamefont {He},
  \citenamefont {Mondaini}, \citenamefont {Luo}, \citenamefont {Wang},\ and\
  \citenamefont {Hu}}]{He2024}%
  \BibitemOpen
  \bibfield  {author} {\bibinfo {author} {\bibfnamefont {W.-X.}\ \bibnamefont
  {He}}, \bibinfo {author} {\bibfnamefont {R.}~\bibnamefont {Mondaini}},
  \bibinfo {author} {\bibfnamefont {H.-G.}\ \bibnamefont {Luo}}, \bibinfo
  {author} {\bibfnamefont {X.}~\bibnamefont {Wang}},\ and\ \bibinfo {author}
  {\bibfnamefont {S.}~\bibnamefont {Hu}},\ }\bibfield  {title} {\bibinfo
  {title} {Phase transitions in the {H}aldane-{H}ubbard model},\ }\href
  {https://doi.org/10.1103/PhysRevB.109.035126} {\bibfield  {journal} {\bibinfo
   {journal} {Phys. Rev. B}\ }\textbf {\bibinfo {volume} {109}},\ \bibinfo
  {pages} {035126} (\bibinfo {year} {2024})}\BibitemShut {NoStop}%
\bibitem [{\citenamefont {Ebrahimkhas}\ \emph {et~al.}(2021)\citenamefont
  {Ebrahimkhas}, \citenamefont {Hafez-Torbati},\ and\ \citenamefont
  {Hofstetter}}]{Ebrahimkhas2021}%
  \BibitemOpen
  \bibfield  {author} {\bibinfo {author} {\bibfnamefont {M.}~\bibnamefont
  {Ebrahimkhas}}, \bibinfo {author} {\bibfnamefont {M.}~\bibnamefont
  {Hafez-Torbati}},\ and\ \bibinfo {author} {\bibfnamefont {W.}~\bibnamefont
  {Hofstetter}},\ }\bibfield  {title} {\bibinfo {title} {Lattice symmetry and
  emergence of antiferromagnetic quantum {H}all states},\ }\href
  {https://doi.org/10.1103/PhysRevB.103.155108} {\bibfield  {journal} {\bibinfo
   {journal} {Phys. Rev. B}\ }\textbf {\bibinfo {volume} {103}},\ \bibinfo
  {pages} {155108} (\bibinfo {year} {2021})}\BibitemShut {NoStop}%
\bibitem [{\citenamefont {Shao}\ \emph {et~al.}(2021)\citenamefont {Shao},
  \citenamefont {Castro}, \citenamefont {Hu},\ and\ \citenamefont
  {Mondaini}}]{Shao2021}%
  \BibitemOpen
  \bibfield  {author} {\bibinfo {author} {\bibfnamefont {C.}~\bibnamefont
  {Shao}}, \bibinfo {author} {\bibfnamefont {E.~V.}\ \bibnamefont {Castro}},
  \bibinfo {author} {\bibfnamefont {S.}~\bibnamefont {Hu}},\ and\ \bibinfo
  {author} {\bibfnamefont {R.}~\bibnamefont {Mondaini}},\ }\bibfield  {title}
  {\bibinfo {title} {Interplay of local order and topology in the extended
  {H}aldane-{H}ubbard model},\ }\href
  {https://doi.org/10.1103/PhysRevB.103.035125} {\bibfield  {journal} {\bibinfo
   {journal} {Phys. Rev. B}\ }\textbf {\bibinfo {volume} {103}},\ \bibinfo
  {pages} {035125} (\bibinfo {year} {2021})}\BibitemShut {NoStop}%
\bibitem [{\citenamefont {Thouless}\ \emph {et~al.}(1982)\citenamefont
  {Thouless}, \citenamefont {Kohmoto}, \citenamefont {Nightingale},\ and\
  \citenamefont {den Nijs}}]{Thouless1982}%
  \BibitemOpen
  \bibfield  {author} {\bibinfo {author} {\bibfnamefont {D.~J.}\ \bibnamefont
  {Thouless}}, \bibinfo {author} {\bibfnamefont {M.}~\bibnamefont {Kohmoto}},
  \bibinfo {author} {\bibfnamefont {M.~P.}\ \bibnamefont {Nightingale}},\ and\
  \bibinfo {author} {\bibfnamefont {M.}~\bibnamefont {den Nijs}},\ }\bibfield
  {title} {\bibinfo {title} {Quantized {Hall Conductance in a Two-Dimensional
  Periodic Potential}},\ }\href {https://doi.org/10.1103/PhysRevLett.49.405}
  {\bibfield  {journal} {\bibinfo  {journal} {Phys. Rev. Lett.}\ }\textbf
  {\bibinfo {volume} {49}},\ \bibinfo {pages} {405–408} (\bibinfo {year}
  {1982})}\BibitemShut {NoStop}%
\bibitem [{\citenamefont {Haldane}(1988)}]{Haldane1988}%
  \BibitemOpen
  \bibfield  {author} {\bibinfo {author} {\bibfnamefont {F.~D.~M.}\
  \bibnamefont {Haldane}},\ }\bibfield  {title} {\bibinfo {title} {Model for a
  {Quantum {H}all Effect without {L}andau Levels: Condensed-Matter Realization
  of the ``Parity Anomaly''}},\ }\href
  {https://doi.org/10.1103/PhysRevLett.61.2015} {\bibfield  {journal} {\bibinfo
   {journal} {Phys. Rev. Lett.}\ }\textbf {\bibinfo {volume} {61}},\ \bibinfo
  {pages} {2015–2018} (\bibinfo {year} {1988})}\BibitemShut {NoStop}%
\bibitem [{\citenamefont {Haldane}(2017)}]{Haldane2017}%
  \BibitemOpen
  \bibfield  {author} {\bibinfo {author} {\bibfnamefont {F.~D.~M.}\
  \bibnamefont {Haldane}},\ }\bibfield  {title} {\bibinfo {title} {Nobel
  {Lecture: Topological} quantum matter},\ }\href
  {https://link.aps.org/doi/10.1103/RevModPhys.89.040502} {\bibfield  {journal}
  {\bibinfo  {journal} {Rev. Mod. Phys.}\ }\textbf {\bibinfo {volume} {89}},\
  \bibinfo {pages} {040502} (\bibinfo {year} {2017})}\BibitemShut {NoStop}%
\bibitem [{\citenamefont {Aidelsburger}\ \emph {et~al.}(2013)\citenamefont
  {Aidelsburger}, \citenamefont {Atala}, \citenamefont {Lohse}, \citenamefont
  {Barreiro}, \citenamefont {Paredes},\ and\ \citenamefont
  {Bloch}}]{Aidelsburger2013}%
  \BibitemOpen
  \bibfield  {author} {\bibinfo {author} {\bibfnamefont {M.}~\bibnamefont
  {Aidelsburger}}, \bibinfo {author} {\bibfnamefont {M.}~\bibnamefont {Atala}},
  \bibinfo {author} {\bibfnamefont {M.}~\bibnamefont {Lohse}}, \bibinfo
  {author} {\bibfnamefont {J.~T.}\ \bibnamefont {Barreiro}}, \bibinfo {author}
  {\bibfnamefont {B.}~\bibnamefont {Paredes}},\ and\ \bibinfo {author}
  {\bibfnamefont {I.}~\bibnamefont {Bloch}},\ }\bibfield  {title} {\bibinfo
  {title} {Realization of the {Hofstadter} {Hamiltonian} with ultracold atoms
  in optical lattices},\ }\href
  {https://doi.org/10.1103/PhysRevLett.111.185301} {\bibfield  {journal}
  {\bibinfo  {journal} {Phys. Rev. Lett.}\ }\textbf {\bibinfo {volume} {111}},\
  \bibinfo {pages} {185301} (\bibinfo {year} {2013})}\BibitemShut {NoStop}%
\bibitem [{\citenamefont {Miyake}\ \emph {et~al.}(2013)\citenamefont {Miyake},
  \citenamefont {Siviloglou}, \citenamefont {Kennedy}, \citenamefont {Burton},\
  and\ \citenamefont {Ketterle}}]{Miyake2013}%
  \BibitemOpen
  \bibfield  {author} {\bibinfo {author} {\bibfnamefont {H.}~\bibnamefont
  {Miyake}}, \bibinfo {author} {\bibfnamefont {G.~A.}\ \bibnamefont
  {Siviloglou}}, \bibinfo {author} {\bibfnamefont {C.~J.}\ \bibnamefont
  {Kennedy}}, \bibinfo {author} {\bibfnamefont {W.~C.}\ \bibnamefont
  {Burton}},\ and\ \bibinfo {author} {\bibfnamefont {W.}~\bibnamefont
  {Ketterle}},\ }\bibfield  {title} {\bibinfo {title} {Realizing the {Harper}
  {Hamiltonian} with {Laser-Assisted Tunneling in Optical Lattices}},\ }\href
  {https://doi.org/10.1103/PhysRevLett.111.185302} {\bibfield  {journal}
  {\bibinfo  {journal} {Phys. Rev. Lett.}\ }\textbf {\bibinfo {volume} {111}},\
  \bibinfo {pages} {185302} (\bibinfo {year} {2013})}\BibitemShut {NoStop}%
\bibitem [{\citenamefont {Jotzu}\ \emph {et~al.}(2014)\citenamefont {Jotzu},
  \citenamefont {Messer}, \citenamefont {Desbuquois}, \citenamefont {Lebrat},
  \citenamefont {Uehlinger}, \citenamefont {Greif},\ and\ \citenamefont
  {Esslinger}}]{Jotzu2014}%
  \BibitemOpen
  \bibfield  {author} {\bibinfo {author} {\bibfnamefont {G.}~\bibnamefont
  {Jotzu}}, \bibinfo {author} {\bibfnamefont {M.}~\bibnamefont {Messer}},
  \bibinfo {author} {\bibfnamefont {R.}~\bibnamefont {Desbuquois}}, \bibinfo
  {author} {\bibfnamefont {M.}~\bibnamefont {Lebrat}}, \bibinfo {author}
  {\bibfnamefont {T.}~\bibnamefont {Uehlinger}}, \bibinfo {author}
  {\bibfnamefont {D.}~\bibnamefont {Greif}},\ and\ \bibinfo {author}
  {\bibfnamefont {T.}~\bibnamefont {Esslinger}},\ }\bibfield  {title} {\bibinfo
  {title} {Experimental realization of the topological {H}aldane model with
  ultracold fermions},\ }\href {https://doi.org/10.1038/nature13915} {\bibfield
   {journal} {\bibinfo  {journal} {Nature}\ }\textbf {\bibinfo {volume}
  {515}},\ \bibinfo {pages} {237} (\bibinfo {year} {2014})}\BibitemShut
  {NoStop}%
\bibitem [{\citenamefont {J\"unemann}\ \emph {et~al.}(2017)\citenamefont
  {J\"unemann}, \citenamefont {Piga}, \citenamefont {Ran}, \citenamefont
  {Lewenstein}, \citenamefont {Rizzi},\ and\ \citenamefont
  {Bermudez}}]{Juenemann2017}%
  \BibitemOpen
  \bibfield  {author} {\bibinfo {author} {\bibfnamefont {J.}~\bibnamefont
  {J\"unemann}}, \bibinfo {author} {\bibfnamefont {A.}~\bibnamefont {Piga}},
  \bibinfo {author} {\bibfnamefont {S.-J.}\ \bibnamefont {Ran}}, \bibinfo
  {author} {\bibfnamefont {M.}~\bibnamefont {Lewenstein}}, \bibinfo {author}
  {\bibfnamefont {M.}~\bibnamefont {Rizzi}},\ and\ \bibinfo {author}
  {\bibfnamefont {A.}~\bibnamefont {Bermudez}},\ }\bibfield  {title} {\bibinfo
  {title} {Exploring interacting topological insulators with ultracold atoms:
  The synthetic {C}reutz-{H}ubbard model},\ }\href
  {https://doi.org/10.1103/PhysRevX.7.031057} {\bibfield  {journal} {\bibinfo
  {journal} {Phys. Rev. X}\ }\textbf {\bibinfo {volume} {7}},\ \bibinfo {pages}
  {031057} (\bibinfo {year} {2017})}\BibitemShut {NoStop}%
\bibitem [{\citenamefont {Hofstetter}\ and\ \citenamefont
  {Qin}(2018)}]{Hofstetter2018}%
  \BibitemOpen
  \bibfield  {author} {\bibinfo {author} {\bibfnamefont {W.}~\bibnamefont
  {Hofstetter}}\ and\ \bibinfo {author} {\bibfnamefont {T.}~\bibnamefont
  {Qin}},\ }\bibfield  {title} {\bibinfo {title} {Quantum simulation of
  strongly correlated condensed matter systems},\ }\href
  {http://stacks.iop.org/0953-4075/51/i=8/a=082001} {\bibfield  {journal}
  {\bibinfo  {journal} {Journal of Physics B: Atomic, Molecular and Optical
  Physics}\ }\textbf {\bibinfo {volume} {51}},\ \bibinfo {pages} {082001}
  (\bibinfo {year} {2018})}\BibitemShut {NoStop}%
\bibitem [{\citenamefont {Rachel}(2018)}]{Rachel2018}%
  \BibitemOpen
  \bibfield  {author} {\bibinfo {author} {\bibfnamefont {S.}~\bibnamefont
  {Rachel}},\ }\bibfield  {title} {\bibinfo {title} {Interacting topological
  insulators: a review},\ }\href {https://doi.org/10.1088/1361-6633/aad6a6}
  {\bibfield  {journal} {\bibinfo  {journal} {Reports on Progress in Physics}\
  }\textbf {\bibinfo {volume} {81}},\ \bibinfo {pages} {116501} (\bibinfo
  {year} {2018})}\BibitemShut {NoStop}%
\bibitem [{\citenamefont {Hatsugai}\ and\ \citenamefont
  {Kohmoto}(1990)}]{Hatsugai1990}%
  \BibitemOpen
  \bibfield  {author} {\bibinfo {author} {\bibfnamefont {Y.}~\bibnamefont
  {Hatsugai}}\ and\ \bibinfo {author} {\bibfnamefont {M.}~\bibnamefont
  {Kohmoto}},\ }\bibfield  {title} {\bibinfo {title} {Energy spectrum and the
  quantum {H}all effect on the square lattice with next-nearest-neighbor
  hopping},\ }\href {https://doi.org/10.1103/PhysRevB.42.8282} {\bibfield
  {journal} {\bibinfo  {journal} {Phys. Rev. B}\ }\textbf {\bibinfo {volume}
  {42}},\ \bibinfo {pages} {8282} (\bibinfo {year} {1990})}\BibitemShut
  {NoStop}%
\bibitem [{\citenamefont {Dalibard}\ \emph {et~al.}(2011)\citenamefont
  {Dalibard}, \citenamefont {Gerbier}, \citenamefont {{Juzeli\ifmmode
  \bar{u}\else ū\fi{}nas}},\ and\ \citenamefont {Öhberg}}]{Dalibard2011}%
  \BibitemOpen
  \bibfield  {author} {\bibinfo {author} {\bibfnamefont {J.}~\bibnamefont
  {Dalibard}}, \bibinfo {author} {\bibfnamefont {F.}~\bibnamefont {Gerbier}},
  \bibinfo {author} {\bibfnamefont {G.}~\bibnamefont {{Juzeli\ifmmode
  \bar{u}\else ū\fi{}nas}}},\ and\ \bibinfo {author} {\bibfnamefont
  {P.}~\bibnamefont {Öhberg}},\ }\bibfield  {title} {\bibinfo {title}
  {Colloquium: Artificial gauge potentials for neutral atoms},\ }\href
  {https://doi.org/10.1103/RevModPhys.83.1523} {\bibfield  {journal} {\bibinfo
  {journal} {Rev. Mod. Phys.}\ }\textbf {\bibinfo {volume} {83}},\ \bibinfo
  {pages} {1523–1543} (\bibinfo {year} {2011})}\BibitemShut {NoStop}%
\bibitem [{\citenamefont {Georges}\ \emph {et~al.}(1996)\citenamefont
  {Georges}, \citenamefont {Kotliar}, \citenamefont {Krauth},\ and\
  \citenamefont {Rozenberg}}]{Georges1996}%
  \BibitemOpen
  \bibfield  {author} {\bibinfo {author} {\bibfnamefont {A.}~\bibnamefont
  {Georges}}, \bibinfo {author} {\bibfnamefont {G.}~\bibnamefont {Kotliar}},
  \bibinfo {author} {\bibfnamefont {W.}~\bibnamefont {Krauth}},\ and\ \bibinfo
  {author} {\bibfnamefont {M.~J.}\ \bibnamefont {Rozenberg}},\ }\bibfield
  {title} {\bibinfo {title} {Dynamical mean-field theory of strongly correlated
  fermion systems and the limit of infinite dimensions},\ }\href
  {https://doi.org/10.1103/RevModPhys.68.13} {\bibfield  {journal} {\bibinfo
  {journal} {Rev. Mod. Phys.}\ }\textbf {\bibinfo {volume} {68}},\ \bibinfo
  {pages} {13} (\bibinfo {year} {1996})}\BibitemShut {NoStop}%
\bibitem [{\citenamefont {Pavarini}\ \emph {et~al.}(2022)\citenamefont
  {Pavarini}, \citenamefont {Koch}, \citenamefont {Lichtenstein},\ and\
  \citenamefont {Vollhardt}}]{Pavarini2022}%
  \BibitemOpen
  \bibinfo {editor} {\bibfnamefont {E.}~\bibnamefont {Pavarini}}, \bibinfo
  {editor} {\bibfnamefont {E.}~\bibnamefont {Koch}}, \bibinfo {editor}
  {\bibfnamefont {A.}~\bibnamefont {Lichtenstein}},\ and\ \bibinfo {editor}
  {\bibfnamefont {D.}~\bibnamefont {Vollhardt}},\ eds.,\ \href
  {https://www.cond-mat.de/events/correl22/manuscripts/correl22.pdf} {\emph
  {\bibinfo {title} {Dynamical mean-field theory of correlated electrons:
  lecture notes of the Autumn School on Correlated Electrons 2022;
  Forschungszentrum J{\"u}lich, 4-7 October 2022}}}\ (\bibinfo {year}
  {2022})\BibitemShut {NoStop}%
\bibitem [{\citenamefont {Budich}\ \emph {et~al.}(2013)\citenamefont {Budich},
  \citenamefont {Trauzettel},\ and\ \citenamefont {Sangiovanni}}]{Budich2013}%
  \BibitemOpen
  \bibfield  {author} {\bibinfo {author} {\bibfnamefont {J.~C.}\ \bibnamefont
  {Budich}}, \bibinfo {author} {\bibfnamefont {B.}~\bibnamefont {Trauzettel}},\
  and\ \bibinfo {author} {\bibfnamefont {G.}~\bibnamefont {Sangiovanni}},\
  }\bibfield  {title} {\bibinfo {title} {Fluctuation-driven topological {Hund}
  insulators},\ }\href {https://doi.org/10.1103/PhysRevB.87.235104} {\bibfield
  {journal} {\bibinfo  {journal} {Phys. Rev. B}\ }\textbf {\bibinfo {volume}
  {87}},\ \bibinfo {pages} {235104} (\bibinfo {year} {2013})}\BibitemShut
  {NoStop}%
\bibitem [{\citenamefont {Amaricci}\ \emph {et~al.}(2018)\citenamefont
  {Amaricci}, \citenamefont {Valli}, \citenamefont {Sangiovanni}, \citenamefont
  {Trauzettel},\ and\ \citenamefont {Capone}}]{Amaricci2018}%
  \BibitemOpen
  \bibfield  {author} {\bibinfo {author} {\bibfnamefont {A.}~\bibnamefont
  {Amaricci}}, \bibinfo {author} {\bibfnamefont {A.}~\bibnamefont {Valli}},
  \bibinfo {author} {\bibfnamefont {G.}~\bibnamefont {Sangiovanni}}, \bibinfo
  {author} {\bibfnamefont {B.}~\bibnamefont {Trauzettel}},\ and\ \bibinfo
  {author} {\bibfnamefont {M.}~\bibnamefont {Capone}},\ }\bibfield  {title}
  {\bibinfo {title} {Coexistence of metallic edge states and antiferromagnetic
  ordering in correlated topological insulators},\ }\href
  {https://doi.org/10.1103/PhysRevB.98.045133} {\bibfield  {journal} {\bibinfo
  {journal} {Phys. Rev. B}\ }\textbf {\bibinfo {volume} {98}},\ \bibinfo
  {pages} {045133} (\bibinfo {year} {2018})}\BibitemShut {NoStop}%
\bibitem [{\citenamefont {Irsigler}\ \emph {et~al.}(2019)\citenamefont
  {Irsigler}, \citenamefont {Zheng},\ and\ \citenamefont
  {Hofstetter}}]{Irsigler2019}%
  \BibitemOpen
  \bibfield  {author} {\bibinfo {author} {\bibfnamefont {B.}~\bibnamefont
  {Irsigler}}, \bibinfo {author} {\bibfnamefont {J.-H.}\ \bibnamefont
  {Zheng}},\ and\ \bibinfo {author} {\bibfnamefont {W.}~\bibnamefont
  {Hofstetter}},\ }\bibfield  {title} {\bibinfo {title} {Interacting
  {Hofstadter Interface}},\ }\href
  {https://doi.org/10.1103/PhysRevLett.122.010406} {\bibfield  {journal}
  {\bibinfo  {journal} {Phys. Rev. Lett.}\ }\textbf {\bibinfo {volume} {122}},\
  \bibinfo {pages} {010406} (\bibinfo {year} {2019})}\BibitemShut {NoStop}%
\bibitem [{\citenamefont {Ebrahimkhas}\ \emph {et~al.}(2022)\citenamefont
  {Ebrahimkhas}, \citenamefont {Uhrig}, \citenamefont {Hofstetter},\ and\
  \citenamefont {Hafez-Torbati}}]{Ebrahimkhas2022}%
  \BibitemOpen
  \bibfield  {author} {\bibinfo {author} {\bibfnamefont {M.}~\bibnamefont
  {Ebrahimkhas}}, \bibinfo {author} {\bibfnamefont {G.~S.}\ \bibnamefont
  {Uhrig}}, \bibinfo {author} {\bibfnamefont {W.}~\bibnamefont {Hofstetter}},\
  and\ \bibinfo {author} {\bibfnamefont {M.}~\bibnamefont {Hafez-Torbati}},\
  }\bibfield  {title} {\bibinfo {title} {Antiferromagnetic {C}hern insulator in
  centrosymmetric systems},\ }\href
  {https://doi.org/10.1103/PhysRevB.106.205107} {\bibfield  {journal} {\bibinfo
   {journal} {Phys. Rev. B}\ }\textbf {\bibinfo {volume} {106}},\ \bibinfo
  {pages} {205107} (\bibinfo {year} {2022})}\BibitemShut {NoStop}%
\bibitem [{\citenamefont {Hafez-Torbati}\ and\ \citenamefont
  {Uhrig}(2024)}]{HafezTorbati2024}%
  \BibitemOpen
  \bibfield  {author} {\bibinfo {author} {\bibfnamefont {M.}~\bibnamefont
  {Hafez-Torbati}}\ and\ \bibinfo {author} {\bibfnamefont {G.~S.}\ \bibnamefont
  {Uhrig}},\ }\bibfield  {title} {\bibinfo {title} {Antiferromagnetic chern
  insulator with large charge gap in heavy transition-metal compounds},\ }\href
  {https://doi.org/10.1038/s41598-024-68044-z} {\bibfield  {journal} {\bibinfo
  {journal} {Scientific Reports}\ }\textbf {\bibinfo {volume} {14}},\ \bibinfo
  {pages} {17168} (\bibinfo {year} {2024})}\BibitemShut {NoStop}%
\bibitem [{\citenamefont {Hafez-Torbati}\ \emph {et~al.}(2020)\citenamefont
  {Hafez-Torbati}, \citenamefont {Zheng}, \citenamefont {Irsigler},\ and\
  \citenamefont {Hofstetter}}]{HafezTorbati2020}%
  \BibitemOpen
  \bibfield  {author} {\bibinfo {author} {\bibfnamefont {M.}~\bibnamefont
  {Hafez-Torbati}}, \bibinfo {author} {\bibfnamefont {J.-H.}\ \bibnamefont
  {Zheng}}, \bibinfo {author} {\bibfnamefont {B.}~\bibnamefont {Irsigler}},\
  and\ \bibinfo {author} {\bibfnamefont {W.}~\bibnamefont {Hofstetter}},\
  }\bibfield  {title} {\bibinfo {title} {Interaction-driven topological phase
  transitions in fermionic {SU(3)} systems},\ }\href
  {https://doi.org/10.1103/PhysRevB.101.245159} {\bibfield  {journal} {\bibinfo
   {journal} {Phys. Rev. B}\ }\textbf {\bibinfo {volume} {101}},\ \bibinfo
  {pages} {245159} (\bibinfo {year} {2020})}\BibitemShut {NoStop}%
\bibitem [{\citenamefont {Potthoff}\ and\ \citenamefont
  {Nolting}(1999)}]{Potthoff1999}%
  \BibitemOpen
  \bibfield  {author} {\bibinfo {author} {\bibfnamefont {M.}~\bibnamefont
  {Potthoff}}\ and\ \bibinfo {author} {\bibfnamefont {W.}~\bibnamefont
  {Nolting}},\ }\bibfield  {title} {\bibinfo {title} {Surface metal-insulator
  transition in the {H}ubbard model},\ }\href
  {https://doi.org/10.1103/PhysRevB.59.2549} {\bibfield  {journal} {\bibinfo
  {journal} {Phys. Rev. B}\ }\textbf {\bibinfo {volume} {59}},\ \bibinfo
  {pages} {2549} (\bibinfo {year} {1999})}\BibitemShut {NoStop}%
\bibitem [{\citenamefont {Snoek}\ \emph {et~al.}(2008)\citenamefont {Snoek},
  \citenamefont {Titvinidze}, \citenamefont {T{\H{o}}ke}, \citenamefont
  {Byczuk},\ and\ \citenamefont {Hofstetter}}]{Snoek2008}%
  \BibitemOpen
  \bibfield  {author} {\bibinfo {author} {\bibfnamefont {M.}~\bibnamefont
  {Snoek}}, \bibinfo {author} {\bibfnamefont {I.}~\bibnamefont {Titvinidze}},
  \bibinfo {author} {\bibfnamefont {C.}~\bibnamefont {T{\H{o}}ke}}, \bibinfo
  {author} {\bibfnamefont {K.}~\bibnamefont {Byczuk}},\ and\ \bibinfo {author}
  {\bibfnamefont {W.}~\bibnamefont {Hofstetter}},\ }\bibfield  {title}
  {\bibinfo {title} {Antiferromagnetic order of strongly interacting fermions
  in a trap: real-space dynamical mean-field analysis},\ }\href
  {https://doi.org/10.1088/1367-2630/10/9/093008} {\bibfield  {journal}
  {\bibinfo  {journal} {New Journal of Physics}\ }\textbf {\bibinfo {volume}
  {10}},\ \bibinfo {pages} {093008} (\bibinfo {year} {2008})}\BibitemShut
  {NoStop}%
\bibitem [{\citenamefont {Hafez-Torbati}\ and\ \citenamefont
  {Hofstetter}(2018)}]{HafezTorbati2018}%
  \BibitemOpen
  \bibfield  {author} {\bibinfo {author} {\bibfnamefont {M.}~\bibnamefont
  {Hafez-Torbati}}\ and\ \bibinfo {author} {\bibfnamefont {W.}~\bibnamefont
  {Hofstetter}},\ }\bibfield  {title} {\bibinfo {title} {Artificial {SU(3)}
  spin-orbit coupling and exotic {Mott} insulators},\ }\href
  {https://doi.org/10.1103/PhysRevB.98.245131} {\bibfield  {journal} {\bibinfo
  {journal} {Phys. Rev. B}\ }\textbf {\bibinfo {volume} {98}},\ \bibinfo
  {pages} {245131} (\bibinfo {year} {2018})}\BibitemShut {NoStop}%
\bibitem [{\citenamefont {Caffarel}\ and\ \citenamefont
  {Krauth}(1994)}]{Caffarel1994}%
  \BibitemOpen
  \bibfield  {author} {\bibinfo {author} {\bibfnamefont {M.}~\bibnamefont
  {Caffarel}}\ and\ \bibinfo {author} {\bibfnamefont {W.}~\bibnamefont
  {Krauth}},\ }\bibfield  {title} {\bibinfo {title} {Exact diagonalization
  approach to correlated fermions in infinite dimensions: Mott transition and
  superconductivity},\ }\href
  {https://link.aps.org/doi/10.1103/PhysRevLett.72.1545} {\bibfield  {journal}
  {\bibinfo  {journal} {Phys. Rev. Lett.}\ }\textbf {\bibinfo {volume} {72}},\
  \bibinfo {pages} {1545} (\bibinfo {year} {1994})}\BibitemShut {NoStop}%
\bibitem [{\citenamefont {Müller-Hartmann}(1989)}]{MuellerHartmann1989}%
  \BibitemOpen
  \bibfield  {author} {\bibinfo {author} {\bibfnamefont {E.}~\bibnamefont
  {Müller-Hartmann}},\ }\bibfield  {title} {\bibinfo {title} {Correlated
  fermions on a lattice in high dimensions},\ }\href
  {https://doi.org/10.1007/BF01311397} {\bibfield  {journal} {\bibinfo
  {journal} {Zeitschrift für Physik B Condensed Matter}\ }\textbf {\bibinfo
  {volume} {74}},\ \bibinfo {pages} {507} (\bibinfo {year} {1989})}\BibitemShut
  {NoStop}%
\bibitem [{\citenamefont {Wang}\ and\ \citenamefont {Zhang}(2012)}]{Wang2012}%
  \BibitemOpen
  \bibfield  {author} {\bibinfo {author} {\bibfnamefont {Z.}~\bibnamefont
  {Wang}}\ and\ \bibinfo {author} {\bibfnamefont {S.-C.}\ \bibnamefont
  {Zhang}},\ }\bibfield  {title} {\bibinfo {title} {Simplified {Topological
  Invariants for Interacting Insulators}},\ }\href
  {https://doi.org/10.1103/PhysRevX.2.031008} {\bibfield  {journal} {\bibinfo
  {journal} {Phys. Rev. X}\ }\textbf {\bibinfo {volume} {2}},\ \bibinfo {pages}
  {031008} (\bibinfo {year} {2012})}\BibitemShut {NoStop}%
\bibitem [{\citenamefont {Wang}\ and\ \citenamefont {Yan}(2013)}]{Wang2013}%
  \BibitemOpen
  \bibfield  {author} {\bibinfo {author} {\bibfnamefont {Z.}~\bibnamefont
  {Wang}}\ and\ \bibinfo {author} {\bibfnamefont {B.}~\bibnamefont {Yan}},\
  }\bibfield  {title} {\bibinfo {title} {Topological {Hamiltonian} as an exact
  tool for topological invariants},\ }\href
  {https://doi.org/10.1088/0953-8984/25/15/155601} {\bibfield  {journal}
  {\bibinfo  {journal} {Journal of Physics: Condensed Matter}\ }\textbf
  {\bibinfo {volume} {25}},\ \bibinfo {pages} {155601} (\bibinfo {year}
  {2013})}\BibitemShut {NoStop}%
\bibitem [{\citenamefont {Gurarie}(2011)}]{Gurarie2011}%
  \BibitemOpen
  \bibfield  {author} {\bibinfo {author} {\bibfnamefont {V.}~\bibnamefont
  {Gurarie}},\ }\bibfield  {title} {\bibinfo {title} {Single-particle {G}reen's
  functions and interacting topological insulators},\ }\href
  {https://doi.org/10.1103/PhysRevB.83.085426} {\bibfield  {journal} {\bibinfo
  {journal} {Phys. Rev. B}\ }\textbf {\bibinfo {volume} {83}},\ \bibinfo
  {pages} {085426} (\bibinfo {year} {2011})}\BibitemShut {NoStop}%
\bibitem [{\citenamefont {Wagner}\ \emph {et~al.}(2023)\citenamefont {Wagner},
  \citenamefont {Crippa}, \citenamefont {Amaricci}, \citenamefont {Hansmann},
  \citenamefont {Klett}, \citenamefont {König}, \citenamefont {Schäfer},
  \citenamefont {Sante}, \citenamefont {Cano}, \citenamefont {Millis},
  \citenamefont {Georges},\ and\ \citenamefont {Sangiovanni}}]{Wagner2023}%
  \BibitemOpen
  \bibfield  {author} {\bibinfo {author} {\bibfnamefont {N.}~\bibnamefont
  {Wagner}}, \bibinfo {author} {\bibfnamefont {L.}~\bibnamefont {Crippa}},
  \bibinfo {author} {\bibfnamefont {A.}~\bibnamefont {Amaricci}}, \bibinfo
  {author} {\bibfnamefont {P.}~\bibnamefont {Hansmann}}, \bibinfo {author}
  {\bibfnamefont {M.}~\bibnamefont {Klett}}, \bibinfo {author} {\bibfnamefont
  {E.~J.}\ \bibnamefont {König}}, \bibinfo {author} {\bibfnamefont
  {T.}~\bibnamefont {Schäfer}}, \bibinfo {author} {\bibfnamefont {D.~D.}\
  \bibnamefont {Sante}}, \bibinfo {author} {\bibfnamefont {J.}~\bibnamefont
  {Cano}}, \bibinfo {author} {\bibfnamefont {A.~J.}\ \bibnamefont {Millis}},
  \bibinfo {author} {\bibfnamefont {A.}~\bibnamefont {Georges}},\ and\ \bibinfo
  {author} {\bibfnamefont {G.}~\bibnamefont {Sangiovanni}},\ }\bibfield
  {title} {\bibinfo {title} {Mott insulators with boundary zeros},\ }\href
  {https://doi.org/10.1038/s41467-023-42773-7} {\bibfield  {journal} {\bibinfo
  {journal} {Nature Communications}\ }\textbf {\bibinfo {volume} {14}},\
  \bibinfo {pages} {7531} (\bibinfo {year} {2023})}\BibitemShut {NoStop}%
\bibitem [{\citenamefont {Blason}\ and\ \citenamefont
  {Fabrizio}(2023)}]{Blason2023}%
  \BibitemOpen
  \bibfield  {author} {\bibinfo {author} {\bibfnamefont {A.}~\bibnamefont
  {Blason}}\ and\ \bibinfo {author} {\bibfnamefont {M.}~\bibnamefont
  {Fabrizio}},\ }\bibfield  {title} {\bibinfo {title} {Unified role of
  {G}reen's function poles and zeros in correlated topological insulators},\
  }\href {https://doi.org/10.1103/PhysRevB.108.125115} {\bibfield  {journal}
  {\bibinfo  {journal} {Phys. Rev. B}\ }\textbf {\bibinfo {volume} {108}},\
  \bibinfo {pages} {125115} (\bibinfo {year} {2023})}\BibitemShut {NoStop}%
\bibitem [{\citenamefont {Potthoff}(2018)}]{Potthoff2018}%
  \BibitemOpen
  \bibfield  {author} {\bibinfo {author} {\bibfnamefont {M.}~\bibnamefont
  {Potthoff}},\ }\bibfield  {title} {\bibinfo {title} {Cluster extensions of
  dynamical mean-field theory, {I}n {E}. {P}avarini, {E}. {K}och, {A}.
  {L}ichtenstein and {D}. {V}ollhardt, eds., {DMFT}: {F}rom {I}nfinite
  {D}imensions to {R}eal {M}aterials},\ }\href
  {https://www.cond-mat.de/events/correl18/manuscripts/potthoff.pdf} {\bibfield
   {journal} {\bibinfo  {journal} {vol. 8, pp. 5.1–5.33. {F}orschungszentrum
  {J}\"ulich}\ } (\bibinfo {year} {2018})}\BibitemShut {NoStop}%
\bibitem [{\citenamefont {Gu}\ \emph {et~al.}(2019)\citenamefont {Gu},
  \citenamefont {Li},\ and\ \citenamefont {Li}}]{Gu2019}%
  \BibitemOpen
  \bibfield  {author} {\bibinfo {author} {\bibfnamefont {Z.-L.}\ \bibnamefont
  {Gu}}, \bibinfo {author} {\bibfnamefont {K.}~\bibnamefont {Li}},\ and\
  \bibinfo {author} {\bibfnamefont {J.-X.}\ \bibnamefont {Li}},\ }\bibfield
  {title} {\bibinfo {title} {Quantum cluster approach to the topological
  invariants in correlated {C}hern insulators},\ }\href
  {https://dx.doi.org/10.1088/1367-2630/ab2a97} {\bibfield  {journal} {\bibinfo
   {journal} {New Journal of Physics}\ }\textbf {\bibinfo {volume} {21}},\
  \bibinfo {pages} {073016} (\bibinfo {year} {2019})}\BibitemShut {NoStop}%
\bibitem [{\citenamefont {Wang}\ \emph {et~al.}(2024)\citenamefont {Wang},
  \citenamefont {Shao}, \citenamefont {Tohyama}, \citenamefont {Luo},\ and\
  \citenamefont {Lu}}]{Wang2024}%
  \BibitemOpen
  \bibfield  {author} {\bibinfo {author} {\bibfnamefont {B.-Q.}\ \bibnamefont
  {Wang}}, \bibinfo {author} {\bibfnamefont {C.}~\bibnamefont {Shao}}, \bibinfo
  {author} {\bibfnamefont {T.}~\bibnamefont {Tohyama}}, \bibinfo {author}
  {\bibfnamefont {H.-G.}\ \bibnamefont {Luo}},\ and\ \bibinfo {author}
  {\bibfnamefont {H.}~\bibnamefont {Lu}},\ }\bibfield  {title} {\bibinfo
  {title} {Topological phase in the extended {H}aldane-{H}ubbard model with
  sublattice-dependent repulsion},\ }\href
  {https://doi.org/10.1103/PhysRevB.110.035107} {\bibfield  {journal} {\bibinfo
   {journal} {Phys. Rev. B}\ }\textbf {\bibinfo {volume} {110}},\ \bibinfo
  {pages} {035107} (\bibinfo {year} {2024})}\BibitemShut {NoStop}%
\bibitem [{\citenamefont {Tran}\ and\ \citenamefont {Tran}(2022)}]{Tran2022}%
  \BibitemOpen
  \bibfield  {author} {\bibinfo {author} {\bibfnamefont {M.-T.}\ \bibnamefont
  {Tran}}\ and\ \bibinfo {author} {\bibfnamefont {T.-M.~T.}\ \bibnamefont
  {Tran}},\ }\bibfield  {title} {\bibinfo {title} {Half-topological state in
  magnetic topological insulators},\ }\href
  {https://doi.org/10.1088/1361-648X/ac699f} {\bibfield  {journal} {\bibinfo
  {journal} {Journal of Physics: Condensed Matter}\ }\textbf {\bibinfo {volume}
  {34}},\ \bibinfo {pages} {275603} (\bibinfo {year} {2022})}\BibitemShut
  {NoStop}%
\bibitem [{\citenamefont {Bloch}\ \emph {et~al.}(2008)\citenamefont {Bloch},
  \citenamefont {Dalibard},\ and\ \citenamefont {Zwerger}}]{Bloch2008}%
  \BibitemOpen
  \bibfield  {author} {\bibinfo {author} {\bibfnamefont {I.}~\bibnamefont
  {Bloch}}, \bibinfo {author} {\bibfnamefont {J.}~\bibnamefont {Dalibard}},\
  and\ \bibinfo {author} {\bibfnamefont {W.}~\bibnamefont {Zwerger}},\
  }\bibfield  {title} {\bibinfo {title} {Many-body physics with ultracold
  gases},\ }\href {https://doi.org/10.1103/RevModPhys.80.885} {\bibfield
  {journal} {\bibinfo  {journal} {Rev. Mod. Phys.}\ }\textbf {\bibinfo {volume}
  {80}},\ \bibinfo {pages} {885} (\bibinfo {year} {2008})}\BibitemShut
  {NoStop}%
\bibitem [{\citenamefont {Aligia}(2004)}]{Aligia2004}%
  \BibitemOpen
  \bibfield  {author} {\bibinfo {author} {\bibfnamefont {A.~A.}\ \bibnamefont
  {Aligia}},\ }\bibfield  {title} {\bibinfo {title} {Charge dynamics in the
  {M}ott insulating phase of the ionic {H}ubbard model},\ }\href
  {https://doi.org/10.1103/PhysRevB.69.041101} {\bibfield  {journal} {\bibinfo
  {journal} {Phys. Rev. B}\ }\textbf {\bibinfo {volume} {69}},\ \bibinfo
  {pages} {041101} (\bibinfo {year} {2004})}\BibitemShut {NoStop}%
\bibitem [{\citenamefont {Shao}\ and\ \citenamefont {Luo}(2024)}]{Shao2024}%
  \BibitemOpen
  \bibfield  {author} {\bibinfo {author} {\bibfnamefont {C.}~\bibnamefont
  {Shao}}\ and\ \bibinfo {author} {\bibfnamefont {H.-G.}\ \bibnamefont {Luo}},\
  }\bibfield  {title} {\bibinfo {title} {Phase diagram of the interacting
  {H}aldane model with spin-dependent sublattice potentials},\ }\href
  {https://doi.org/10.1103/PhysRevB.109.235101} {\bibfield  {journal} {\bibinfo
   {journal} {Phys. Rev. B}\ }\textbf {\bibinfo {volume} {109}},\ \bibinfo
  {pages} {235101} (\bibinfo {year} {2024})}\BibitemShut {NoStop}%
\end{thebibliography}
%

\end{document}